\documentclass[toc]{proar}

\title{Observables in Strongly Coupled Anisotropic Theories}

\ShortTitle{Strongly Coupled Anisotropic Plasmas}

\author{Dimitrios Giataganas${}^{1,2}$\\
1)Department of Engineering Sciences, University of Patras,
26110 Patras, Greece\\
2)National Institute for Theoretical Physics, School of Physics
and Centre for Theoretical Physics,
University of the Witwatersrand, Wits, 2050, South Africa\\
E-mail: \email{dgiataganas@upatras.gr}}

\abstract{We review certain anisotropic gauge/gravity dualities, focusing more on a theory with space dependent axion term. Then we discuss and also present some new results for several observables: the static potential and force, the imaginary part of the static potential, the quark dipole in the plasma wind, the drag force and diffusion time, the jet quenching of heavy and light quarks, the energy loss of rotating quarks, the photon production and finally the violation of the holographic viscosity over entropy bound. The corresponding weakly coupled results are also discussed. Finally we investigate the bounds of the parameters of the current strongly coupled anisotropic theories attempting to match them with the observed quark-gluon plasma and report the problems appear.}

\FullConference{}

\begin{document}

\tableofcontents
\section{Introduction}
In the ultra-relativistic heavy-ion collision experiments
there is an extensive study on the properties of the Quark Gluon Plasma
(QGP). By now there is a very little doubt that the QGP is a strongly coupled fluid \cite{RHIC}. Moreover, there is a strong belief that the QGP is far from equilibrium after its creation and for a short period, where anisotropies occur both in momentum and coordinate space.

\begin{figure*}[!ht]
\centerline{\includegraphics[width=70mm]{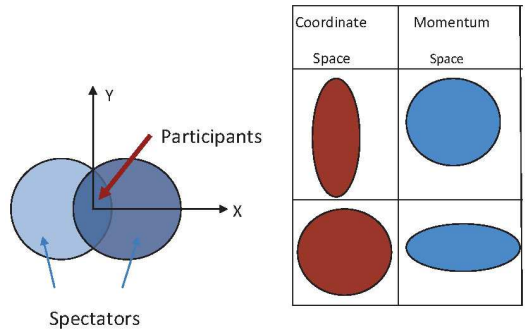}}
\caption{{ \small The transverse plane $xy$ to the collision axis. The participants contribute in the collisions where the different pressure gradients along the short and the long axis 
generate anisotropic momentum distributions.
\label{fig:figee1}
}}
\end{figure*}

The QGP goes through different phases in a short period of time. It can be thought as a small simulation of the big bang with certain basic differences being for example the relative number of baryons present at the process and the time scales involved. The QGP is forming at $\tau\simeq 0.1 fm$  after the collisions. Then due to pressure anisotropies follows the interesting non-equilibrium period, where momentum and coordinate space anisotropies occur. This lasts approximately for $0.1 fm\lesssim \tau\lesssim .3-2fm$, where the high bound on the time has not been specified yet very accurately, but the thermalization and isotropization process of the plasma is currently under intensive studies. The  times of order $\tau\lesssim 2 fm$ are predicted using conformal viscus hydrodynamics, but the values depend strongly on the initial conditions and the details of the plasma hadronization. The gauge/gravity duality models suggest lower thermalization times $\sim 0.3 fm$ \cite{s.3}. At the anisotropic stage several parameters that describe the plasma properties have been introduced. For example, the elliptic flow parameter
\be
v_2=\frac{\vev{p_x^2-p_y^2}}{\vev{p_x^2+p_y^2}}~,
\ee
where $p_{x,y}$ are the momenta of the particles in the transverse space of the collision in Figure \ref{fig:figee1} can be measured experimentally though the particle distributions and can be calculated theoretically with hydrodynamic simulations, predicting low thermalization times.
At RHIC energies the elliptic flow is described relatively well by models which apply the hydrodynamics at $\tau\lesssim 1 fm$.
Then for $.3-2fm\lesssim \tau\lesssim 15 fm$ the equilibrium phase of the plasma follows. This phase is under better control and has been studied extensively. Then for a time period of $3 fm$, $15fm\lesssim \t\lesssim 18 fm$ a hot hadron gas is formed and for later times $\tau\gtrsim 18 fm$ the freezout starts to occur.

Here we are mainly interested in the anisotropic phase of the plasma.  One anisotropy appears because the plasma expands mainly along the collision axis.  Studying the anisotropies at the weak coupling limit, color plasma instabilities occur which are responsible for the short isotropization time in the QGP. The non-Abelian plasma instabilities lead also to turbulent behavior \cite{Arnold:2005ef}. A hard-thermal-loop effective theory of plasma in the equilibrium has been generalized for a non-equilibrium anisotropic plasmas in \cite{Mrowczynski:2004kv}. In the weakly coupled theory have been studied several observables, some of them are the heavy quark potential and the quarkonium properties \cite{qqaniso,0901.1998,qa2,qa1}, the jet quenching \cite{jetqa1,jetqa2} and the photon and dilepton emission \cite{photon2,photon}. Anisotropic expanding plasmas have also been considered in \cite{expanding} and which lead to an anomalous viscosity with low value \cite{Asakawa:2006tc}.  Other anisotropic hydrodynamical models that interpolate between an anisotropic initial state and a state that is described by the perfect fluid hydrodynamics have been developed  in \cite{1007.0889,Martinez:2010sd} and \cite{Ryblewski:2012rr}.

Taking into account that the QGP is a strongly coupled fluid the perturbative methods of Quantum Chromodynamics (QCD) are in general not appropriate for describing it. An alternative approach could be by using the Lattice field theory, for example as in \cite{lns}, but the progress is difficult since real time phenomena in finite temperature need to be studied. Another promising approach to study these phenomena, at least in a qualitative level, is by using the methods of the gauge/gravity dualities \cite{adscft}. The theories used in the dualities to study QGP properties have in general different characteristics than QCD. Nevertheless, in high temperature many of these theories have common characteristics and it is believed that there exist a certain level of universality in their results. The most known achievement of the holographic studies of the plasma, is the shear viscosity over entropy density calculation where a very low value for the fraction is predicted \cite{etas}.  The QGP plasma at equilibrium has been studied extensively using the gauge/gravity correspondence \cite{review1}. It is natural and very interesting to attempt to extend these studies to the non-equilibrium phase.  In this paper we discuss some recent progress to this direction.

In heavy-ion collisions there are several collective phenomena generated by the anisotropies, ie. longitudinal flow where the collective motion is along the beam, radial flow where the velocity of the particles have spherical symmetry, elliptic flow where the particles preferably move along certain azimuthal angles as described before. We are mostly interested here to the longitudinal expansion of the plasma along the collision axis. A way to partly isolate this anisotropy is to think the colliding nuclei having infinite transverse area, or more realistically to participate only in completely central collisions. Using this setup after the formation time, a rapid longitudinal expansion along the beam line occurs. The pressures along the longitudinal and transverse direction follow the inequality $P_L<P_T$ and the particle momenta $\vev{p_L^2}<\vev{p_T^2}$ in the local rest frame.

To resemble such an anisotropic plasma in the gravity dual theory one would aim to have a time dependent anisotropy. However, even the models with a fixed anisotropy provide important information for the anisotropy effects to the plasma. Here we use such static anisotropic models and we review and present some new results on several observables in the anisotropic gauge/gravity dualities.

Our paper has the following structure. In section 2 we review two recent anisotropic backgrounds, one background obtained by a space dependent  $\theta$ term deformation of the $\cN=4$ sYM \cite{mateosaniso1} and the other which is singular derived from a stationary anisotropic energy momentum tensor \cite{janikaniso1}. We discuss the dual field theory of the first background and focus on their most interesting properties. In section 3 we discuss some weak coupling models and establish a connection between the parameters appearing in the strongly coupled background and the weak coupling models through the pressure formulas. Having ready the basic setup we move to study several observables in sections 4 and 5: the static potential and force, the imaginary part of the static potential, the quark dipole in the plasma wind, the drag force and diffusion time, the jet quenching of heavy and light quarks, the energy loss of rotating quarks, the photon production and finally the violation of the holographic viscosity over entropy bound.   At several points the results of the two models are compared to each other, as well as to the weak coupling results. In section 6, we show that more "quantitative" predictions of the observed expanding QGP are not possible using these particular dualities due to certain bounds of their parameters.

\section{Gauge Theory and Dual Gravity Setup}

In this section we review two anisotropic backgrounds we use to calculate observables in the anisotropic QGP. We focus mostly on the first anisotropic background for various reasons which are analyzed below, the most important being the singularity which appears in the background 2.

\subsection{Background 1 \cite{mateosaniso1}}

The anisotropic background of this section is a solution to the type IIB supergravity equations and is a top-down construction. In the field theory side it can be though as a deformed version of the $\cN=4$ finite temperature sYM. The additional deformation term in the action is a $\theta$ parameter term depending on the anisotropic direction \cite{takayanagi}
\be\label{cs}
\delta S\sim\int \a x_3 Tr F\wedge F~,
\ee
where $x_3$ is the spatial anisotropic coordinate. If one choose to compactify the $x_3$-direction on a circle, the remaining three dimensional theory contains a normal Chern-Simons term which couples to the charge density $\a$. If additionally the usual antiperiodic conditions will be imposed for the fermions around the circle the theory flows to a Chern-Simons theory in IR and applications to holographic Quantum Hall effect can be studied \cite{QHE}. In principle space-dependent $\theta$ deformations do not have to break the supersymmetry \cite{cssusy}, although in our case they do.

In the gravity dual side the $\theta$ angle is related to the axion of the type IIB supergravity through the complexified coupling constant
\be
\tau=\frac{\theta\prt{x_3}}{2\pi}+\frac{4 \pi i}{g_{YM}^2}\Leftrightarrow\chi\prt{x_3}+i e^{-\phi}
\ee
and therefore an axion with space dependence will be present in the anisotropic background.
\begin{figure*}[!ht]
\centerline{\includegraphics[width=73mm]{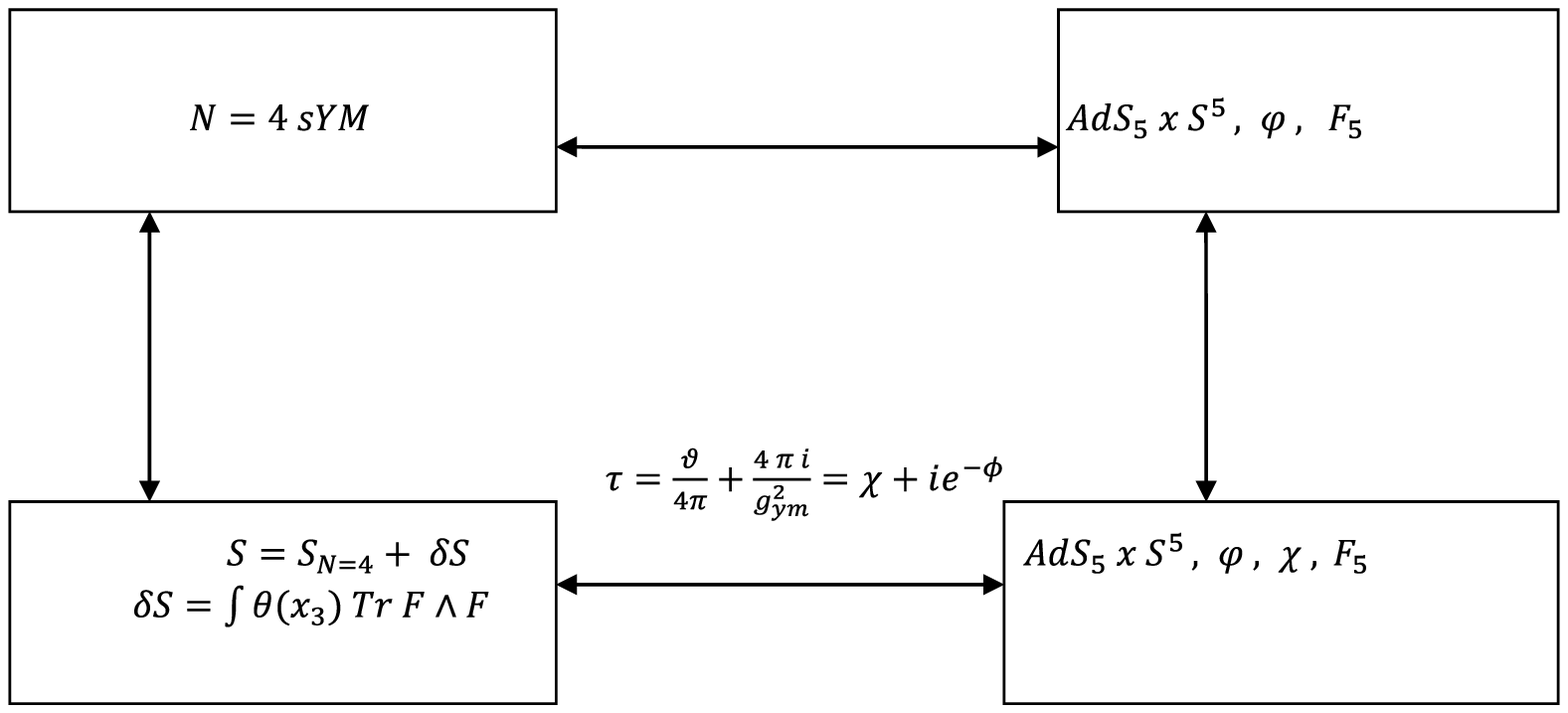}}
\caption{{ \small The deformation of the $\cN=4$ sYM with the space dependent $\theta$ term, in field theory and the gravity side.}}\vspace{-.5cm}
\label{fig:fige2}.
\end{figure*}
The presence of the axion is related with the presence of D7 branes. It can be understood through the RR forms, where the RR 8-form $C_8$ gives $dC_8\sim\star d\chi$ where the $\star$ is the Hodge dual in ten dimensions. Therefore the directions where $C_8$ lies are completely specified as $C_{x_0 x_1 x_2~S^5}$, where with $S^5$ are denoted the five coordinates of the sphere. As a result the $D7$ branes wrap the $S^5$ and extend along the three space time directions $x_0,x_1,x_2$, while can be though as smeared along the anisotropic direction.

\centerline{\includegraphics[width=85mm]{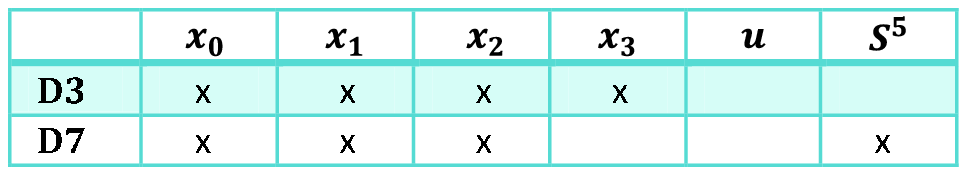}}
As a larger number of $D7$ branes is taken, they start to backreact on the geometry anisotropically and they deform it in a way that the final geometry has a broken rotational invariance. Notice that the $D7$ branes do not extent along the holographic radial direction, therefore do not touch the boundary and do not add new degrees of freedom to the theory. This is their essential difference with the flavor branes.

The ansatz taken for the axion is $\chi=\a x_3$, where the proportionality constant turns out to be related to the density $n_{D7}$ of the smeared branes along the anisotropic direction as $\a=\lambda n_{D7}/4\pi N_c$. The geometry of the supergravity solution has a singularity in the IR which is hidden behind the horizon. Moreover in the boundary it becomes asymptotically AdS while in the deep IR approaches a Lifshitz-like solution found in \cite{takayanagi}, where for example the invariant scaling takes the form $\prt{x_0,x_{1,2},x_3,u}\rightarrow\prt{\lambda x_0,\lambda x_{1,2},\lambda^{3/2}x_3,\lambda u}$. So the solution can be viewed as a renormalization group flow from an isotropic UV point to and anisotropic IR. It seems very possible to exist other solutions with  similar characteristics.

In the string frame the background is given by
\bea
&&
ds^2 =
 \frac{1}{u^2}\left( -\cF \cB\, dx_0^2+dx_1^2+dx_2^2+\cH dx_3^2 +\frac{ du^2}{\cF}\right)+ {\cal Z} \, d\Omega^2_{S^5}\,.
 \label{metric} \\
&&  \chi = \a x_3, \qquad \phi=\phi(u) \,,\qquad
\label{chi}
\eea
where $\phi$ is the dilaton, $\chi$ is the axion, and $\a$ is the anisotropic parameter measured in units of inverse length. The functions
$\cF, \cB,\cZ$ depend on the holographic coordinate $u$ and their form will be given later. The $x_3$ direction is the anisotropic one and will be referred as a longitudinal to anisotropy, where the other two on the transverse plane which has the rotational symmetry will be referred as the transverse directions. The background has a RR-five from proportional to the volume of the sphere
\be
F_5=4\prt{\Omega_{S^5}+\star\Omega_{S^5}}
\ee
and has no $B$-field.
The analytical form of the functions can be found when the scale anisotropy over temperature is small enough, $\a/T\ll1$. Here we consider the expansions of the fields up to second order in $\a/T$ around a black D3-brane solution
\bea
\cF(u) &=& 1 - \frac{u^4}{u_h^4} + \a^2 \cF_2 (u)  +\mathcal{O}(\a^4)\\
\cB(u) &=& 1 + \a^2 \cB_2 (u) +\mathcal{O}(\a^4)\,, \\
\cH(u)&=&e^{-\phi(u)},\quad\cZ(u)x=e^{\frac{\phi(u)}{2}},\quad\mbox{where}\quad \phi(u) =  \a^2 \phi_2 (u)  +\mathcal{O}(\a^4)\,.
\label{smallae}
\eea
The exact form of the $\cF_2 (u), \cB_2 (u), \phi_2 (u)$ function can be found by solving the Einstein equations with asymptotic $AdS$ boundary conditions and requiring $\cF_2$ to vanish at the horizon, giving
\bea
\cF_2(u)&=& \frac{1}{24 u_h^2}\left[8 u^2( u_h^2-u^2)-10 u^4\log 2 +(3 u_h^4+7u^4)\log\left(1+\frac{u^2}{u_h^2}\right)\right]\,,\cr
B_2(u)&=& -\frac{u_h^2}{24}\left[\frac{10 u^2}{u_h^2+u^2} +\log\left(1+\frac{u^2}{u_h^2}\right)\right] \,,\cr
\phi_2(u) &=& -\frac{u_h^2}{4}\log\left(1+\frac{u^2}{u_h^2}\right)\,.
\eea
The temperature of the solution can be found following the usual procedure as
\be\label{tuh1}
T=-\frac{\partial_u\cF \sqrt{\cB}}{4\pi}\bigg|_{u=u_h}=\frac{1}{\pi u_h}+\a^2 u_h \frac{5 \log2-2}{48 \pi} +\mathcal{O}(\a^4)
\ee
and the $u_h$ parameter can be traded with the more physical parameter, the temperature by
\be\label{uht1}
u_h=\frac{1}{\pi T}+\a^2 \frac{5 \log2-2}{48 \pi^3 T^3}+\mathcal{O}(\a^4)~.
\ee
The energy and the pressures can be found from the expectation values of the stress tensor, with
$P_\perp:=\vev{T_{11}}=\vev{T_{22}}=P_{x_1 x_2}$ the pressures along the $x_1$ or $x_2$ directions and $P_\parallel:=\vev{T_{33}}=P_{x_3}$ the pressure along the anisotropic direction
\bea
E&=& \frac{3\pi^2  N_c^2  T^4}{8}+ \a^2
\frac{ N_c^2  T^2}{32}+{O}(\a^4)\,,\cr
P_{x_1 x_2}&=& \frac{\pi^2  N_c^2  T^4}{8}+
\a^2\frac{ N_c^2  T^2}{32}+{O}(\a^4)\,,\cr
P_{x_3}&=& \frac{\pi^2  N_c^2  T^4}{8}-
\a^2\frac{ N_c^2  T^2}{32}+{O}(\a^4)~. \label{pxyz1}
\eea
At low anisotropies the pressure of the plasma along the anisotropic direction is always lower than the one in the other two directions
\be\label{pressaniso1}
P_{x_3}<P_{x_1 x_2}~,
\ee
and this inequality specifies the region we are mostly interested in the
QGP. The entropy density reads
\be
s \propto\frac{\pi^2  N_c^2   T^3}{2}+\a^2 \frac{ N_c^2  T}{16}+\mathcal{O}(\a^4)\,.
\label{entra22}
\ee
Notice that all the expressions for zero anisotropies approach the isotropic D3-brane solution smoothly.

For larger anisotropies the solution can be found numerically. However after a critical value of $\a/T$ the inequality \eq{pressaniso1} gets reversed. Also the solution at this region has different scalings in entropy density where
\be
s\propto N_c^2 \a^{1/3}T^{8/3}~.
\ee
Notice that the phase diagram of a non-interacting version of this theory has been considered in \cite{Gynther:2012mw}.

\subsection{Background 2  \cite{janikaniso1}}

The background under discussion in this section is a stationary anisotropic with energy-momentum tensor
\be\label{t1}
\vev{T_{\mu\nu}}=diag\left(\epsilon,P_L,P_T,P_T\right),\qquad \vev{T_{\mu}^{\nu}}=0~,
\ee
and $\e=P_L+2P_T$. The  most generic metric having these symmetries
is
\be\label{j1}
ds^2=\frac{1}{u^2}\left(-a\prt{u}dt^2+b\prt{u}dx_{3,L}^2+c\prt{u}\prt{dx_{1,T}^2+dx_{2,T}^2}+du^2\right) ~,
\ee
where $a\prt{u},b\prt{u},c\prt{u}$ functions are specified by the boundary conditions.
To see how,  we use the fact that in Fefferman-Graham coordinates the metric of an asymptotically AdS space can be written in general as
\be
ds^2=\left(g_{\mu\nu}\left(x,u\right)+du^2\right) \frac{1}{u^2}~,
\ee
where $u$ is the holographic coordinate and the boundary of the space is at $u=0$. Near the boundary the metric satisfies
\be
g_{\mu\nu}\left(x,u\right)=\eta_{\m\n}+u^4\gamma_{\m\n}^\prt{4}\left(x,u\right)+\cO\prt{u^6}
\ee
and is a solution of the Einstein equation $R_{\m\n}\prt{x,u}=-4 g_{\m\n}\prt{x,u}$. The $\g^\prt{4}_{\m\n}$ is given by the expectation value of the energy momentum tensor \cite{emT}
\be
\vev{T_{\m\n}}=\frac{N_c^2}{2\pi^2}\gamma_{\m\n}^\prt{4}
\ee
and the metric for a certain energy momentum tensor is chosen by the conditions that no singularities appear in the resulting bulk geometry \cite{geomT}. The functions of the metric \eq{j1} can be specified as
\bea
a(u)&=&(1+A^2u^4)^{\frac{1}{2}-\frac{1}{4}\sqrt{36-2B^2}}   (1-A^2u^4)^{\frac{1}{2}+\frac{1}{4}\sqrt{36-2B^2}}~,\\
b(u)&=&(1+A^2u^4)^{\frac{1}{2}-\frac{B}{3}+\frac{1}{12}\sqrt{36-2B^2}}  (1-A^2u^4)^{\frac{1}{2}+\frac{B}{3}-\frac{1}{12}\sqrt{36-2B^2}}~,\\
c(u)&=&(1+A^2u^4)^{\frac{1}{2}+\frac{B}{6}+\frac{1}{12}\sqrt{36-2B^2}}   (1-A^2u^4)^{\frac{1}{2}-\frac{B}{6}-\frac{1}{12}\sqrt{36-2B^2}}~,
\eea
where for $u\rightarrow 0$ it approaches the AdS space and
with the two free parameters $A$ and $B$ appearing in the energy density and pressure relations
\bea
\e=\frac{1}{2} A^2 \sqrt{36- 2 B^2}~,\quad
P_L =\frac{1}{6} A^2 \sqrt{36-2 B^2} -\frac{2}{3} A^2 B~,\quad
P_T =\frac{1}{6} A^2 \sqrt{36-2 B^2} +\frac{1}{3} A^2 B~,
\eea
normalized appropriately. Notice that $B=0$, results equal pressures and the solution reduces to a static AdS black hole. For $B<0$ the plasma is prolate and for the special value $B=-\sqrt{6}\Rightarrow P_T=0$. For $B>0$ the plasma is oblate and for $B=\sqrt{2}\Rightarrow P_L=0$.

However when the anisotropy is present the metric has a naked singularity in the bulk. Nevertheless the singularity is in a sense mild, since for the scalar wave equation, ingoing and outgoing boundary conditions can be naturally defined.
\footnote{Other holographic models that have been constructed for anisotropic fluids are in \cite{otheraniso}.
Quite different anisotropies may be sourced by external magnetic field as for example in \cite{Ammon:2012qs}.}
\section{Relation of the Background Parameters to the Anisotropic QGP and Connection to the Weakly Coupled Anisotropy Parameters}

To make the connection of the background anisotropy and the plasma we need to consider a kinematic example where the accelerated beams of nucleons collide along the anisotropic $x_3$ direction, the beam axis direction, where the plasma initially expands rapidly.

\begin{figure}
\centerline{\includegraphics[height=60mm]{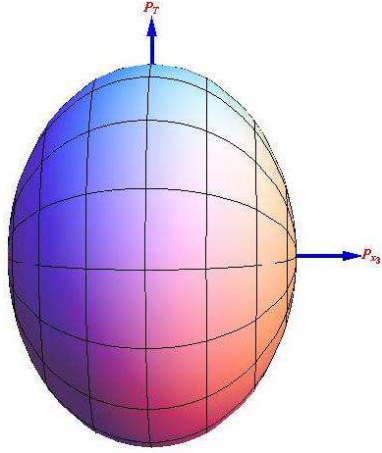}}
\caption{\small{Anisotropy in the momentum space for $\xi>0$.}}\label{fig:figob1}
\end{figure}

A usual method to represent the geometry of the anisotropic plasma in the weakly coupled theories is to consider its distribution function $f\prt{t,\textbf{x},\textbf{p}}$ homogeneous in position space but anisotropic in the momentum space. It takes the form
\cite{anisofunction1}
\be\label{faniso}
f_{aniso}(\textbf{p})=c_{norm}(\xi) f_{iso}(\sqrt{\textbf{p}^2+\xi (\textbf{p}\cdot \textbf{n})^2})~,
\ee
where the vector $\textbf{n}=(0,0,1)$ is the unit vector chosen along the anisotropic direction and the parameter $\xi$ is a measure of anisotropy and $c_{norm}$ is the normalization constant.
The function deforms the spherical momentum distribution since along the anisotropic direction the momentum is rescaled. For $\xi>0$ the distribution is contracted in the anisotropic direction and it is the configuration we are interested mostly depicted in Figure \ref{fig:figob1}. For $-1<\xi<0$ the distribution is stretched along the anisotropic direction. The anisotropic QGP created in the heavy ion collisions corresponds to $\xi>0$, and this is why this region interests us most.
The anisotropy parameter can be expressed in terms of the average particle momenta along the two transverse and longitudinal directions as
\be\label{xidefine2}
\xi=\frac{\vev{p_T^2}}{2 \vev{p_L^2}}-1~,\quad\mbox{where}\quad\textbf{p}_T=\textbf{p}-\textbf{n}(\textbf{p}\cdot \textbf{n})~,\quad p_L=\textbf{p}\cdot \textbf{n}~,
\ee
where the role of the anisotropic parameter in the plasma becomes clear. It should be noted here that the normalization constant $c_{norm}$ plays an important role to the subsequent results. By modifying the argument of the function only, the anisotropic plasma can be obtained from the isotropic one by removing or adding particles with large momentum along the anisotropic direction as can also be seen from \eq{xidefine2}. Therefore, the density of the plasma changes along the different directions and it is natural to expect that certain observables will be modified due to this fact, apart from the anisotropic effects. The constant however can be fixed by requiring the density of the hard particles to be equal along the isotropic and anisotropic directions
\be
\int f_{iso}\prt{p}=c_{norm}(\xi)\int f_{iso}(\sqrt{\textbf{p}^2+\xi (\textbf{p}\cdot \textbf{n})^2})\Rightarrow c_{norm}(\xi)=\sqrt{1+\xi}~.
\ee
Another way to specify the constant is by requiring constant energy density which gives
\be
c_{norm}(\xi)=\prt{\frac{1}{2}\prt{1+|\xi|}^{-1}+ |\xi|^{-1/2}\arctan\sqrt{|\xi|}}~.
\ee
The distribution function then is used to study the weakly coupled anisotropic plasmas, where for example a splitting in the soft modes that carry momenta of order $g_{YM} T$ and the hard modes that carry momenta of order $T$ can be imposed. The latter ones are the particles that at least should have the anisotropic phase space distribution function.

It is possible to establish a naive quantitative connection of the anisotropic parameter $\xi$ with the parameters $\a$ and $B$ appear in the gravity backgrounds 1 and 2 respectively. This can be done through the pressure inequalities of the systems. We introduce the parameter $\D$ which measures the degree of pressure anisotropy through the different directions as
\be
\D:=\frac{P_T}{P_L}-1=\frac{P_{x_1 x_2}}{P_{x_3}}-1~.
\ee
The pressures along the different directions are written as
\be
P_i=T_{ii}=c_{norm}(\xi) \int \frac{d^3\textbf{p} p_{i}^2}{\prt{2 \pi}^3|\textbf{p}|}f_{iso}(\sqrt{\textbf{p}^2+\xi (\textbf{p}\cdot \textbf{n})^2})~,
\ee
which give
\be
P_T=\frac{3 c_{norm}}{4 \xi}\prt{1+\prt{\xi-1}\frac{\arctan\sqrt{\xi}}{\sqrt{\xi}}}P_{iso}\quad\mbox{and}\quad P_L=\frac{3 c_{norm}}{2 \xi}\prt{\frac{\arctan\sqrt{\xi}}{\sqrt{\xi}}-\frac{\xi}{1+\xi}}P_{iso}
\ee
and therefore $\D$ can be expressed in terms of $\xi$ as \cite{0902.3834}
\be\label{dxi1}
\D=\frac{1}{2}(\xi-3)+\xi\left(\left(1+\xi\right)\frac{\arctan\sqrt{\xi}}
{\sqrt{\xi}}-1 \right)^{-1}~.
\ee
To obtain analytical results of $\xi$ in terms of $\D$ we need to take the limit of small or large anisotropies. For small anisotropy
\be\label{dx1b}
\D=\frac{4}{5}\xi +\cO(\xi^2)~,
\ee
where we expect to correspond to small values of the anisotropic parameter $\a$.

Using \eq{pxyz1} we get the relevant equation of \eq{dxi1} for the gravity background as
\be
\label{da1b}
\D=\frac{\a^2}{2 \pi^2 T^2}~,
\ee
where it can be seen that small $\D$ corresponds to small $\a/T$ and therefore
we can use the approximation \eq{dx1b} to get
\be\label{xia1}
\xi\backsimeq \frac{5 \a^2}{8 \pi^2 T^2}~,
\ee
equation that is valid for $\a\ll T\Leftrightarrow\D\ll 1\Leftrightarrow \xi \ll 1$. We should note here that the equation \eq{xia1} which provides a naive quantitative connection between the anisotropic parameters in supergravity backgrounds and the anisotropic momentum distribution functions considered in several field theory models, it is obtained only through the pressure anisotropies of the models. The way that the pressure anisotropy is generated did not play a direct role.

Thinking similarly for the singular background 2, by fixing the number density and setting $A=1$ on the supergravity solution, the relation between $B$ and $\xi$ reads \cite{Rebhan:2011ke}
\be
\frac{\sqrt{36-2 B^2}+2B}{\sqrt{36-2 B^2}-4B}=\frac{\xi-1}{2}+\frac{\xi}{\prt{\xi+1}\xi^{-1/2}\arctan\xi^{1/2}-1}~,
\ee
and for small $\xi$ becomes
\be
\xi=\frac{5}{4}B+\cO\prt{B^2}~.
\ee
A general comment is that the comparison of the strongly coupled results to the weakly coupled should be interpreted with caution. First of all the source of the anisotropies are different: in the weak coupling case it is due to the distribution of the particles in momentum phase, while in the strong coupling case it is due to an external source in the dual theory of the background 1. Moreover, the strongly coupled models do not take into account the dynamical flavor degrees of freedom which might affect the results qualitative. Nevertheless, it is possible that the observables might depend stronger on the resulted anisotropy related to the final pressure inequality of the plasma and not the way it is generated and therefore the comparisons may be interesting. It should be also clear that weak and strong coupling observables even in the case of same theories do not have to necessary match.

\section{Observables in Strongly Coupled Anisotropic Models}

Here we study various observables found in the anisotropic plasmas. We will focus more on observables in the $\theta$ dependent anisotropic background, but in several cases we will discuss results on the other background too. A basic reference we follow for most of the observables and methodology used is the reference \cite{Giataganas_aniso}, while for observables which are not covered in this reference, we mention the relevant references in the section where the observable is discussed. Some new results and discussions are also presented.

\subsection{Q\={Q} Static Potential and Static Force in the Anisotropic Plasma}

In this section we study the static potential and force in the finite temperature anisotropic plasma. This was firstly done in \cite{Giataganas_aniso} and then other interesting aspects were also analyzed in \cite{1205.4684,1208.2672,Chakraborty:2012dt}.
In order to focus on the main text mostly on the anisotropic results, we
use the Appendix \ref{app:QQ} to briefly present the gravity calculations of the Q\={Q} Wilson loop.

Notice that the anisotropic axion-deformed background satisfies the UV divergencies cancelation conditions for Wilson Loops derived in \cite{giataganasUV} and it is expected that the mass subtraction scheme will give consistent results with the Legendre transform for the UV divergencies cancelation. Moreover, the form of the potential is similar to the isotropic theory including the turning point behavior \cite{Reyy}, where the lowest energy branch is the physical, while the non-physical and unstable \cite{sfetsos} the rest of it, where the string is already broken.

In the anisotropic plasma we have two different perpendicular static potentials, one in the transverse plane and one in the anisotropic direction. We will find and compare them each other and with the isotropic result.
The corresponding world-sheets can be expressed by
\bea
x_0=\t\qquad\mbox{and}\qquad x_p=\sigma,
\qquad u=u(\s)~,\\
\mbox{where}\quad x_p=x_1=:x_\perp\quad\mbox{or}\quad x_p=x_3=:x_\parallel~.
\eea
The static potential in anisotropic background has the same form with the finite temperature isotropic background. For fixed small anisotropy $\a/T$, the potentials satisfy in absolute value
\be\label{vrelation1}
V_{\parallel}<V_{\perp}<V_{iso}~.
\ee
As a consequence of this relation is that the critical length is reduced in presence of anisotropy as
\be\label{lrelation1}
L_{c\parallel}<L_{c\perp}<L_{c~iso}~.
\ee
It is interesting to note that the degree of modification due to anisotropy in the results we obtain is in some sense expected, since the anisotropic direction depends more heavily on the anisotropic parameter $\a$ and therefore the results along this direction are expected to be modified stronger. Of course this is not a generic rule since the final behavior specified from the way that the observable depend on the metric elements of the particular directions.

\begin{figure*}[!ht]
\begin{minipage}[ht]{0.5\textwidth}
\begin{flushleft}
\centerline{\includegraphics[width=70mm]{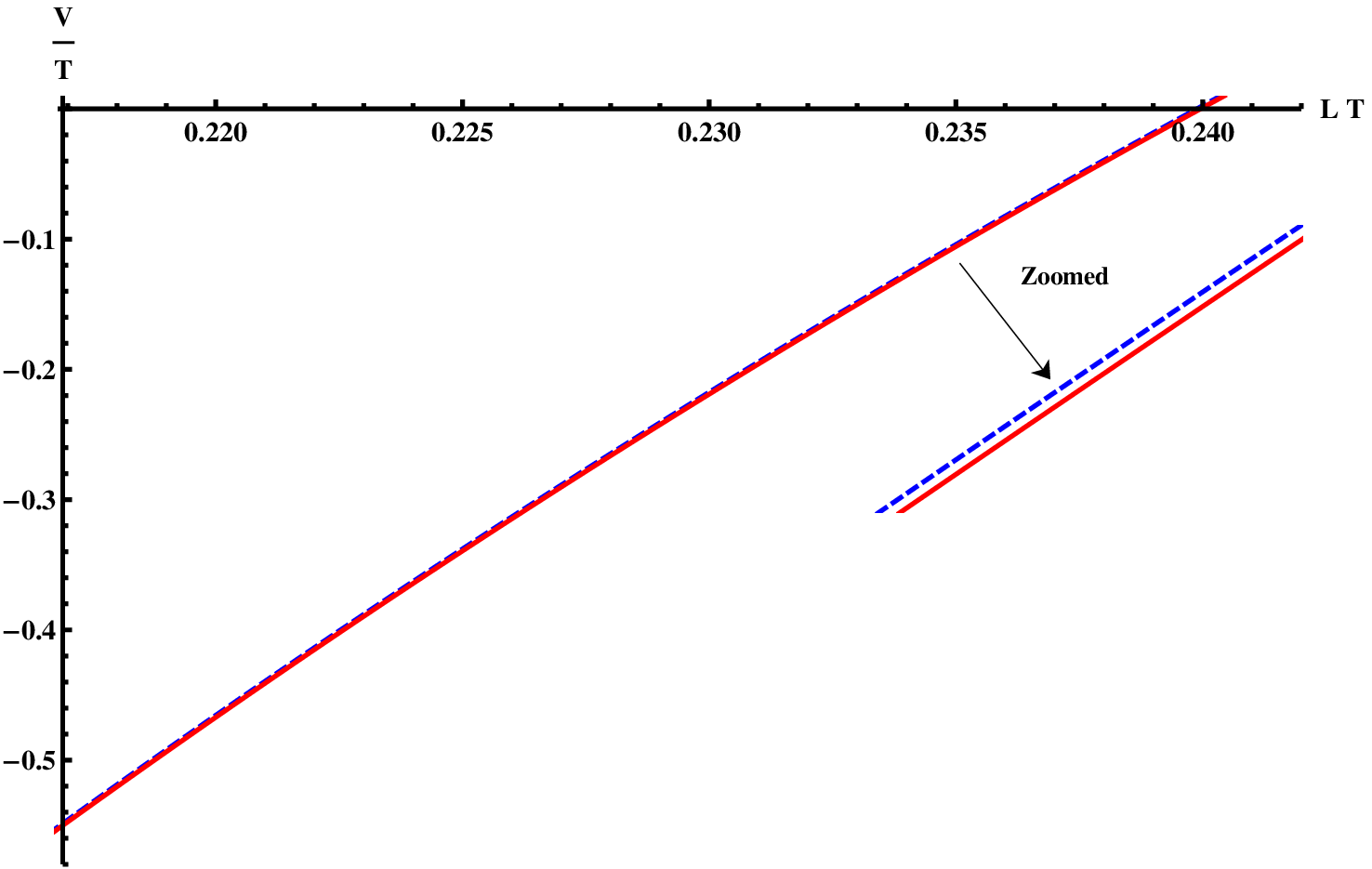}}
\caption{{\small The static potential vs $L T$, close to $L_c$ for Q\={Q} pairs aligned along the anisotropic and the transverse direction. The relation $|V_\parallel|<|V_\perp|$ and $L_{c\parallel}<L_{c\perp}$ is obvious. Settings: blue dotdashed line-$V_\parallel$, red solid line-$V_\perp$ and $T=3$, $\a=0.3 T$.
}}\label{fig:a1aa1}
\end{flushleft}
\end{minipage}
\hspace{0.3cm}
\begin{minipage}[ht]{0.5\textwidth}
\begin{flushleft}
\centerline{\includegraphics[width=70mm]{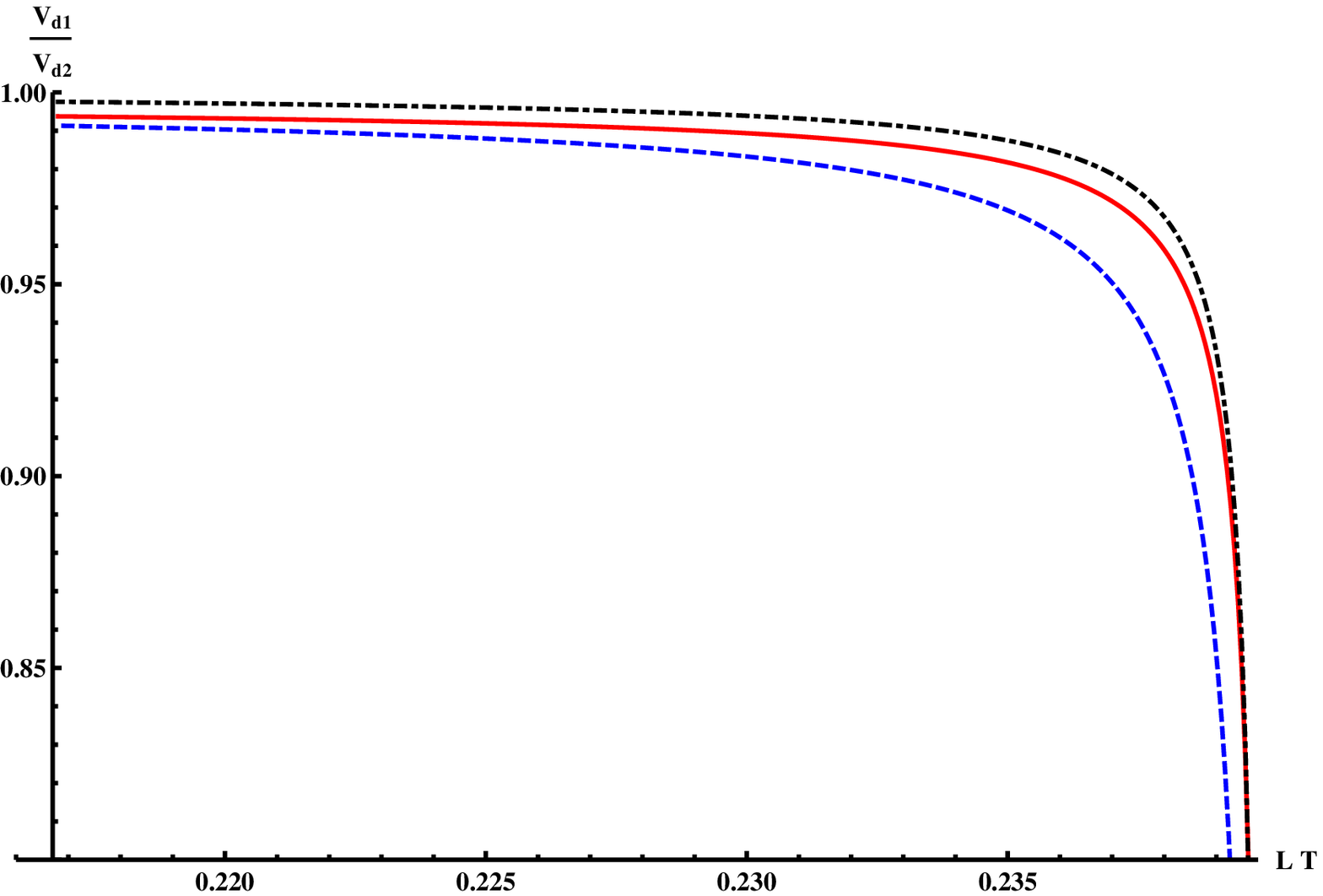}}
\caption{\small{The ratios of the static potentials for pairs aligned along different directions. Settings:$V_\parallel/V_\perp$-black dot-dashed color, $V_\parallel/V_{iso}$-blue dashed line, $V_\perp/V_{iso}$-solid red line and $T=3$, $\a=0.35 T$.}}\label{fig:a2a}
\end{flushleft}
\end{minipage}
\end{figure*}

In addition, increase of the anisotropy leads to decrease of the absolute value of the static potential compared to the undeformed case. Consequence of this is the decrease of the critical length
\be
\a\nearrow~\Rightarrow V_{\parallel,\perp}\searrow~\Rightarrow L_{c~\parallel,\perp} \searrow~.
\ee
These results are presented in Figures \ref{fig:a1aa1} and \ref{fig:a2a}.

The static potential usually has a constant term which has a not clear physical interpretation. The comparison of the quark interactions in presence of anisotropy can be made more manifest if we consider the static forces, where due to the derivative  $F_{Q\bar{Q}}=\frac{\partial V}{\partial L}$ taken on the static potential the constant term disappears \footnote{In \cite{Giataganas:2011nz} a more detailed discussion on static force in AdS/CFT is given.}. We find that the static forces in presence of anisotropy are decreased. The difference along the two direction is much smaller than the static potential and it seems to follow the order
\be\label{forceqq1}
F_{Q\bar{Q},\perp}<F_{Q\bar{Q},\parallel}<F_{iso}~.
\ee
There are several results of the static potential in the weak coupling regime which do not match exactly with the findings of the strong coupling models we have used. In \cite{qqaniso} it has been found enhanced potential in presence of anisotropy with higher magnitude along the transverse direction.  There it was also found that as the separation dipole length is reduced, the difference on the potentials along the different directions get reduced \footnote{This is expected to happen, since in general the anisotropic effects should get weaker for very small distances. In our background as the dipole length is reduced the string worldsheet gets closer to the asymptotically $AdS_5$ boundary, producing  results closer to the isotropic theory.}. The differences between our results and the results of the weak coupling models on the strength of the static potentials might be generated from the different values of the constant in the static potential, and therefore would be more physical to compare the static forces.
It should be noted that the two models have several important differences apart from the main difference that we are comparing results in different regimes of the coupling constants. The comparison happens in order to mention the similarities and differences in the observables but with no expectation to obtain matchings.

A short summary of the strongly coupled results so far is that the presence of anisotropy leads to decrease of the static potential, and further decrease happens for pairs aligned along the anisotropic direction \eq{vrelation1}. Increase of the anisotropy leads to the further decrease of the static potential. The static force also decreases in presence of anisotropy \eq{forceqq1}. The critical lengths for increasing anisotropy, reduce according to \eq{lrelation1}. All the comparisons here are made with equal temperatures.

Although most of the information of the anisotropy can be extracted from the results along the anisotropic and transverse directions, one can also generalize the calculation to obtain results with dependence on the angle between the transverse plane and the anisotropic direction. This may be done by introducing the ansatz
\be
x_{1}\rightarrow \sin\theta x_{1}~,\quad x_3\rightarrow \cos\theta x_{3}~.
\ee
A similar analysis of the Wilson loop in the radial gauge has been considered in \cite{1208.2672}, where the observable results are angle dependent. The calculation methodology is similar to the Appendix \ref{app:QQ}. For $\theta=0$ the pair is aligned along the anisotropic direction and for $\theta=\pi/2$ the pair is aligned along the transverse plane. In these cases the analysis reproduce the findings \cite{Giataganas_aniso} described in detail above. For all other angles the results go smoothly in a monotonic way from $\theta=0$ to the ones for $\theta=\pi/2$. On top on that one can compare the results in different comparison schemes. In the constant entropy density comparison scheme, the screening length along the anisotropic direction decreases with the anisotropy while in the transverse plane it increases. The corresponding static potentials are modified accordingly.

In the background \eq{j1} the critical length depends on the sign of the anisotropic parameter $B$, since oblate and prolate geometries there are possible. For the prolate geometries, the heavy quark results are similar in the two backgrounds. For oblate geometry in \eq{j1} the dipole along the transverse direction has a smaller dissociation distance compared to the isotropic case, while the one along the transverse direction has a larger distance when keeping constant the energy density \cite{1205.4684}.

\subsection{Imaginary potential}

The quarkonium physics in the thermal medium it is expected to depend not only on the real part of the potential, ie. the binding energy, but also to an additional effect that is taking place. The quarkonium may be thought as a quantum mechanical bound state, but with a partly lost coherence due to heat bath energy of the gluons, effect which corresponds to the appearance of an imaginary part of the potential. Such imaginary part in the potential affects the quarkonium decay process. There are various methods to calculate the imaginary potential, most of them inequivalent and in some cases also the presence of anisotropy is studied \cite{imaginary}. Moreover, there are some proposed ways to extend the computation in the strongly coupled regime using the gauge/gravity dualities eg. \cite{Noronha:2009da}.

One can use fluctuations of the worldsheet close to the horizon, large enough to generate the imaginary part of the potential. When this is done, the on-shell the Nambu-Goto action generates a complex part. For a generic anisotropic background this method gives \cite{Giataganas_im}
\be
Im V_{Q\bar{Q}}=
\frac{1}{2\sqrt{2}\alpha'}\left[\frac{f_0}{|f_0'|}-\frac{|f_0'|}{2f_0''}
\right]~\sqrt{g_0}~.\label{im-potential2}
\ee
with $f(u):=-G_{00} G_{pp}~,\quad g(u):=-G_{00} G_{uu}~$ and $f_0=f(u_0)$ and so on. It should be noted that this method generates an unexpected turning point of the imaginary potential, similar to the one of the real part, possibly indicating limitations of the method used to obtain the imaginary part in full generality.
Applying the formula to the low temperature anisotropic background we obtain an increase of the imaginary part in presence of anisotropies when the temperature is kept fixed
\be\label{imin1}
|ImV_{iso}|<|ImV_{\perp}|<|ImV_{\parallel}|~.
\ee
When the entropy density is kept fixed, the above results are modified according to
\be\label{imin12b}
|ImV_{\perp}|<|ImV_{iso}|<|ImV_{\parallel}|~.
\ee
To calculate the thermal width of quarkonia we extrapolate to a straight line the imaginary part and compute the expectation value $\Gamma=-<\psi|ImV_{Q\bar{Q}}|\psi>$, with $|\psi>$ being the ground state of the unperturbed static potential depending on the anisotropic parameter. In the usual $\cN=4$ sYM it is the Coulomb potential. Calculation of the thermal width leads to decrease of the quantity in presence of the anisotropy with the following order
\be
\Gamma_{\perp}<\Gamma_{\parallel}<\Gamma_{iso}~.
\ee
In \cite{1210.7512} the imaginary potentials and the suppression of the bottomonium states in presence of anisotropy is reviewed in weakly coupled plasma and compared with good agreement with the available data.

\subsection{Q\={Q} in the Plasma Wind}

In this section the dipole pair is considered in a moving anisotropic plasma with velocity $v$. The most generic case can be studied by considering the velocity on the $x_1 x_3$ plane, since the $x_1 x_2$ is rotationally invariant. Therefore an $SO\prt{2}$ rotation matrix is applied to $M_1=R_1 \tilde{M_1}$ as a coordinate transformation, where $M_1=\prt{x_1,x_3}^T$ with the usual rotation matrix
\begin{displaymath}
R_1=\left(\begin{array}{c c}
\sin\theta & \cos\theta \\
\cos\theta & -\sin\theta
\end{array}\right)
\end{displaymath}
and then a boost along the $x_3$ direction, $\tilde{M_2}=L_2 \tilde{\tilde{M_2}}$ with $\tilde{M_2}=\prt{\tilde{x_0},\tilde{x_3}}^T$, is applied with matrix
\be
L_2=\left(\begin{array}{c c}
\gamma  & -\gamma v \\
-\gamma v & \gamma
\end{array}\right)
\ee
with $\gamma=\prt{1-v^2}^{-1/2}$. The new metric after the coordinate transformations has non-diagonal elements and depends explicitly on the angle $\theta$ and the velocity $v$. It takes a form \cite{1208.2672}
\be
ds^2=\frac{1}{u^2}\prt{G_{00} dx_0^2+G_{11} dx_1^2+G_{22} dx_2^2+G_{33} dx_3^2+G_{13} dx_1 dx_3+G_{01} dx_0 dx_1+G_{03} dx_0 dx_3+\frac{du^2}{\cF}}~.
\ee
Since the rotation and the boost has been performed we are looking at the more generic position of the pair. The ansatz for the string may be chosen in the radial gauge while allowing the string to extend in both $x_1$ and $x_3$ coordinates
\be
x_0=\t~,\quad u=\s~,\quad x_1=x_1\prt{u}~,\quad x_3=x_3\prt{u}~.
\ee
The process of solving the system goes along the lines of Appendix \ref{app:QQ}. It is interesting that the screening length for ultrarelativistic velocities outside the transverse plane scales as  $L_\parallel\simeq \prt{1-v^2}^{1/2}$ \cite{1208.2672} in contrast  to the usual isotropic $1/4$ power \cite{wind1,wind2}. In transverse plane however  the usual power is restored as $L_\perp\simeq \prt{1-v^2}^{1/4}$. The modification of the power compared to the isotropic case is not unexpected, since the radial scalings in the metric elements are different. Similar are the reasons to the deviation of the universal result of the viscosity over entropy ratio as we will mention later.

\subsection{Drag Force on the Heavy Quarks in the Anisotropic Plasma}\label{section:drag}
In this section we review the results of \cite{Giataganas_aniso} for the drag force and the diffusion time. We consider a heavy quark moving along the anisotropic direction and then a quark moving in the transverse plane in an infinite volume of plasma, and we calculate the force need to be imposed in the quark in order to move with constant velocity $v$. The velocity $v$ has a lower bound in order to have the moving quark well above the subthermal velocities and an upper bound in order to have a dominant deceleration forces due to dragging and not radiation.

\begin{figure*}[!ht]
\begin{minipage}[ht]{0.5\textwidth}
\begin{flushleft}
\centerline{\includegraphics[width=70mm]{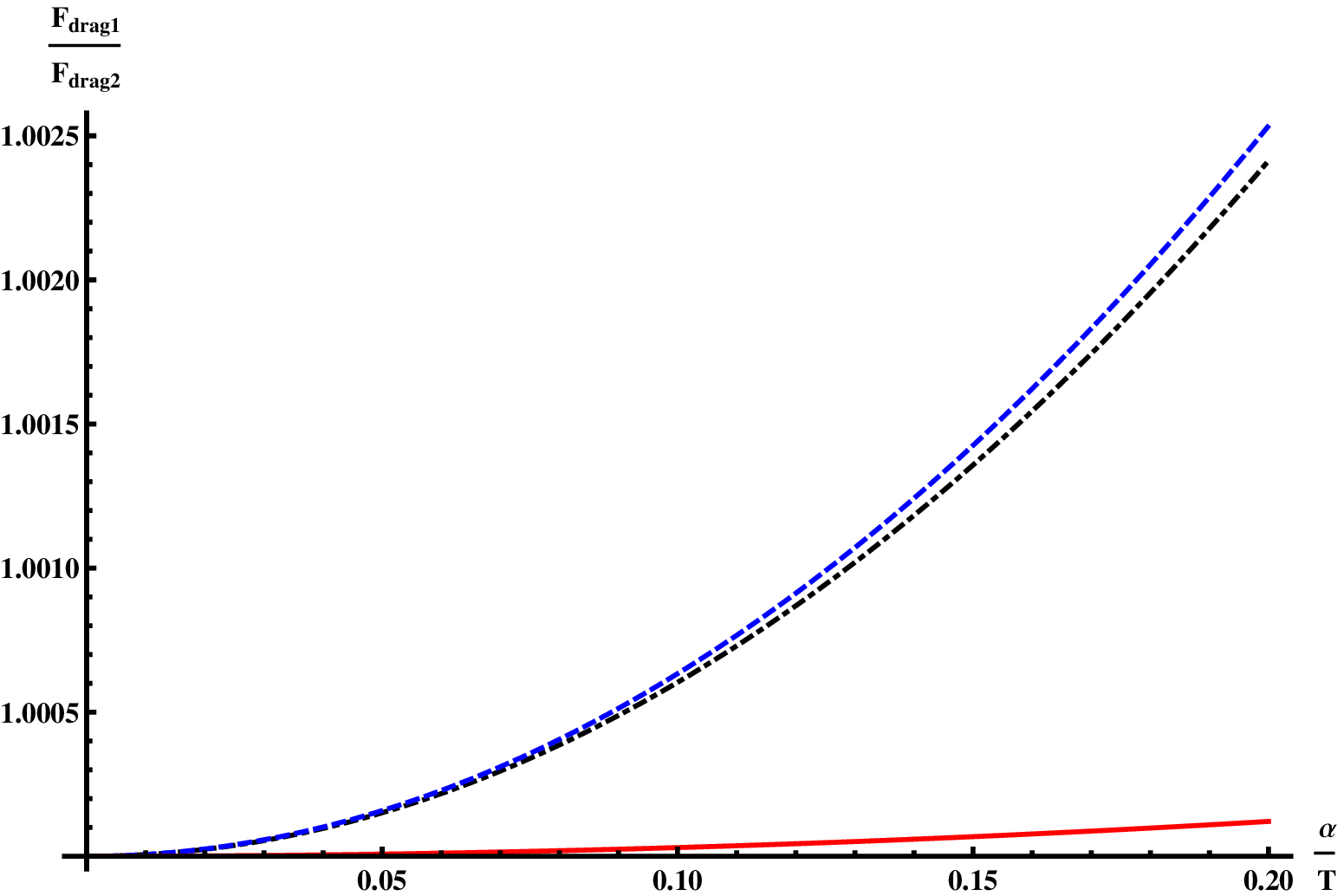}}
\caption{{\small The drag force with the anisotropic parameter $\a/T$, for $v\simeq 0.98$. The ratios in this plot are greater than one, but if we plot for $v\lesssim 0.909$ the $F_{drag,\perp}/F_{drag,iso}<1$, and is reduced as the anisotropy increases.
Settings: black dot-dashed line-$F_{drag,\parallel}/F_{drag,\perp}$, blue dashed line-$F_{drag,\parallel}/F_{drag,iso}$, red solid line-$F_{drag,\perp}/F_{drag,iso}$ and $T=1$.
}}\label{fig:dd1}
\end{flushleft}
\end{minipage}
\hspace{0.3cm}
\begin{minipage}[ht]{0.5\textwidth}
\begin{flushleft}
\centerline{\includegraphics[width=70mm]{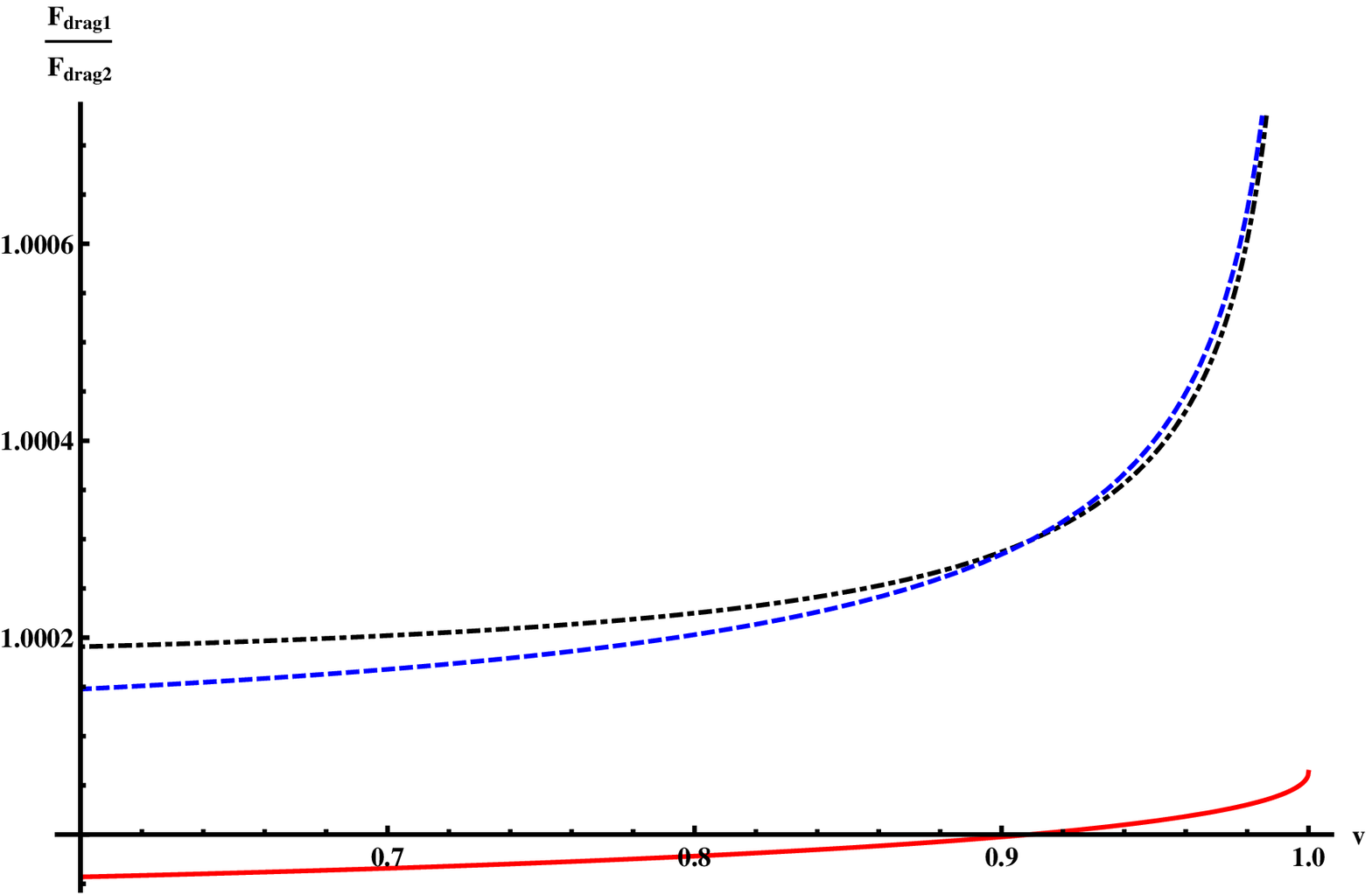}}
\caption{{\small The drag force dependence on the quark velocity $v$. Notice the smaller than the unit values of the ratio $F_{drag,\perp}/F_{drag,iso}$ for $v\lesssim 0.909$.
Settings: $\a=0.1$, $T=1$ and the rest as in Figure 6. \vspace{1.2cm} }}\label{fig:dd2}
\end{flushleft}
\end{minipage}
\end{figure*}

The motion of the string can be formulated as
\be\label{dragansatz}
x_0=\t, \qquad u=\s, \qquad x_p= v \t +f(u)~.
\ee
The generic dual calculation is presented briefly in Appendix \ref{app:drag}. It is interesting to note that since there are no integrations involved here the results are expected to be analytical. The holographic distance $u_0$ can be calculated for the different directions solving the equations \eq{dragu0a1} with the functions
\be
u_{0\parallel}=u_{01\parallel}+\a^2 u_{02\parallel}~, \qquad u_{0 \perp}=u_{01\perp}+\a^2 u_{02\perp}~,
\ee
where exact expressions of $u_{02\parallel,\perp}$ are given in \cite{Giataganas_aniso}. Plugging these solutions to \eq{dragfinala1} we obtain
\bea\label{fpar1}
\frac{F_{drag,\parallel}}{\sqrt{\lambda}}&=&-\frac{\pi T^2 v}{2\sqrt{1-v^2}}\\\nn
&-&\a^2 \frac{v}{48 \pi}\left( \frac{v^2}{ \left(1-v^2\right) \left(1+\sqrt{1-v^2}\right)} +\frac{2(1-v^2)+ \left(1+v^2\right)  \log\left(1+\sqrt{1-v^2}\right)}{\left(1-v^2\right)^{3/2} }\right)~,\\\label{fper1}
\frac{F_{drag,\perp}}{\sqrt{\lambda}}&=&-\frac{\pi T^2 v}{2\sqrt{1-v^2}}\\\nn
&-&\a^2 \frac{v}{48 \pi} \left( \frac{ v^2}{ \left(1-v^2\right) \left(1+\sqrt{1-v^2}\right)}+ \frac{2(1-v^2) - \left(5-4 v^2\right) \log\left(1+\sqrt{1-v^2}\right)}{\left(1-v^2\right)^{3/2}}\right)~,
\eea
where the anisotropic modification is independent of the temperature as expected from the dimensional analysis. Along the anisotropic direction the new term is always negative indicating an increase in the drag force in the anisotropic plasma. In the transverse plane there is a critical value for the velocity $v_c$, and when $v\lesssim 0.909$ the corresponding drag force $F_{drag,\perp}$ is decreased, while in the complementary region we have enhancement. Comparison of the drag forces along the different directions show that the forces required along the anisotropy are always stronger than the ones in the transverse plane
\be\label{fdppb1}
\frac{F_{drag,\parallel}}{F_{drag,\perp}}=1+\a^2 \frac{ \left(2-v^2\right) \log\left(1+\sqrt{1-v^2}\right)}{8 \pi ^2 T^2 \left(1-v^2\right)}~.
\ee
In Figures \ref{fig:dd1} and \ref{fig:dd2} we plot the fraction of the forces depending on the anisotropy and the velocity respectively, and we show schematically the analytic results.

In \cite{1202.3696} the drag force coefficient was calculated for angles between the transverse and longitudinal directions using ansatz similar to \eq{dragansatz} but with
\be
x_1\prt{\t,v}\rightarrow \prt{v \t +f_1(u)}\sin\theta~,\quad x_3\prt{\t,v}\rightarrow \prt{v \t +f_3(u)}\cos\theta~.
\ee
The force goes smoothly and in a monotonic way from the parallel \eq{fpar1} to transverse direction results \eq{fper1}. Using the entropy density comparison scheme, the results turn out to be qualitative similar, something that is expected from the independence of the temperature of the modification terms in the drag force in equations \eq{fpar1} and \eq{fper1}. For larger anisotropies the results are also qualitatively similar, while the velocity $v_c$ approaches the unit.

\begin{figure*}[!ht]
\begin{minipage}[ht]{0.5\textwidth}
\begin{flushleft}
\centerline{\includegraphics[width=70mm]{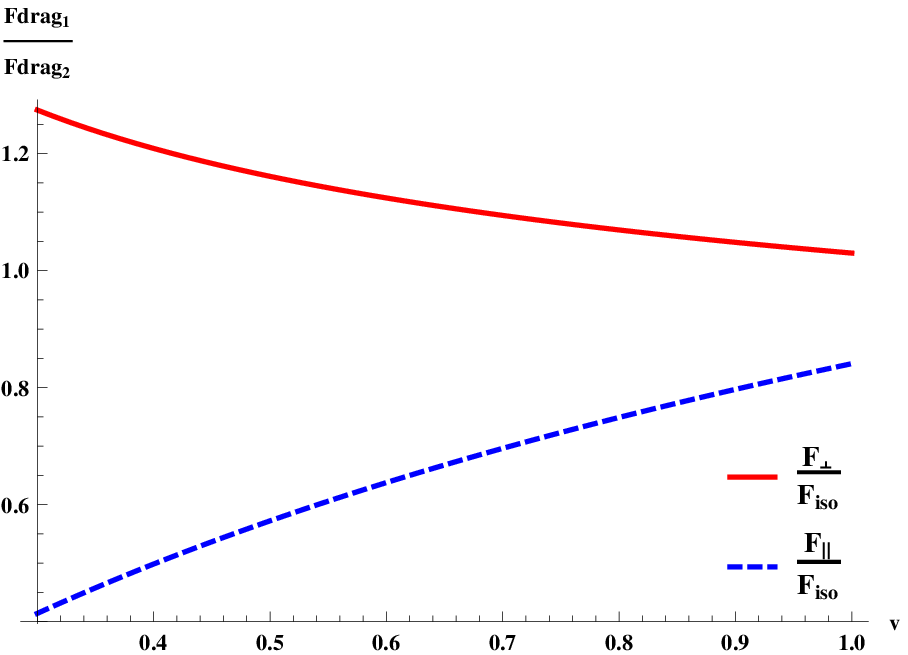}}
\caption{{\small The drag force for the oblate $\prt{B=\sqrt{2}}$ singular geometry depending on the velocity. The drag force along the parallel direction is decreased while in the transverse direction is enhanced.
Plot Settings: as in Figure 6 .}}\label{fig:f1}
\end{flushleft}
\end{minipage}
\hspace{0.3cm}
\begin{minipage}[ht]{0.5\textwidth}
\begin{flushleft}
\centerline{\includegraphics[width=70mm]{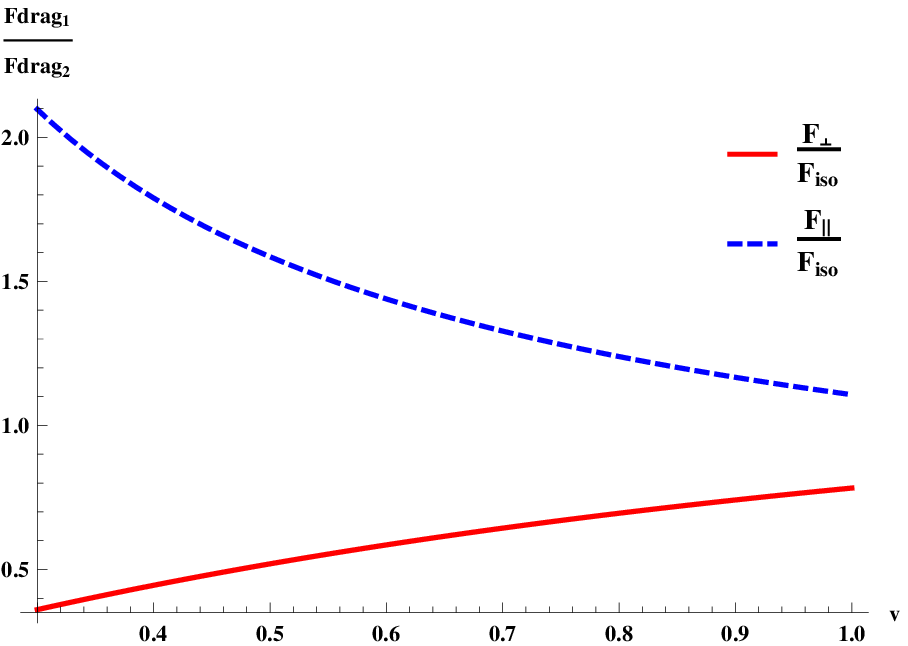}}
\caption{{\small Same plot as on the left but for a prolate geometry $\prt{B=-\sqrt{6}}$. The relation between the drag forces are inverted for the prolate geometry compared to the oblate one. Plot Settings: as in Figure 6. \vspace{0.3cm} }}\label{fig:f2}
\end{flushleft}
\end{minipage}
\end{figure*}

In the singular geometry 2 we present the new results of the drag force
for the prolate and oblate geometries for fixed temperature. The relations of the forces along the different directions get reversed for oblate and prolate geometries as can be seen in Figures \ref{fig:f1} and \ref{fig:f2}.

The \textbf{diffusion time} is the time for which the initial momentum is reduced by $e^{-1}$ factor. It is defined as the inverse of the drag coefficient $n_D$
\be
\frac{dp}{dt}=-n_D \Rightarrow
\t_{D,\parallel,\perp}=-\frac{1}{F_{drag,\parallel,\perp}}
\frac{M_Q v}{\sqrt{1-v^2}}~.
\ee
The mass of the quark $M_Q$ may be fixed by the infinite string with regulator such that the $M_Q\prt{T=0}$ is equal to the physical mass of the quark, and is independent of the direction in the plasma, although it depends on the anisotropic parameter. Therefore, we can take the ratio of the  diffusion times along the two directions in the anisotropic plasma without worrying for the mass deformation due to anisotropy since it will drop out. On top of that, by supposing that the $M_Q$ modification does not affect significantly the results, all the ratios of diffusion times including the isotropic time, can be taken to be independent of the mass of the quark. Of course in a more simplified approach the $M_Q$ can be thought directly as the mass of the heavy probe quark and constant and in that case again the ratios are independent of it. This arguing leads to inverted diffusion time inequalities compared to the drag forces.

To summarise briefly the results of this section, we have found that along the anisotropic direction the drag force is always increased while in the transverse direction, for velocities less than the critical the drag force is decreased:
\bea\nn
&&F_{drag,\parallel}>F_{drag,iso}\quad \mbox{and}\quad F_{drag,\parallel}>F_{drag,\perp}~,\\\nn
&&F_{drag,\perp}>F_{drag,iso}\quad \mbox{for}\quad v>v_c,\quad \mbox{while for}\quad v>v_c~,\quad F_{drag,\perp}<F_{drag,iso} .
\eea
While when using the singular geometry 2 we have found that oblate geometries give $F_{drag,\parallel}$ $<F_{drag,iso}<F_{drag,\perp}$ and prolate geometries give opposite results. \footnote{The drag force in an anisotropic background with Kerr-AdS black hole has been obtained in \cite{schalm}.}

\subsection{Jet Quenching in the Anisotropic Plasma}

The jet quenching is the ratio of mean transverse momentum obtained by the hard parton in the anisotropic medium over the distance it has traveled. The gravity dual calculation can be performed by a world-sheet ending on an orthogonal Wilson loop along two light-like lines with long lengths $L_-$. The long edges of the loop are related to the partons moving at relativistic velocities while the small lengths $L_k$ are related to the transverse momentum broadening of the radiated gluons.

The generic calculation of the jet quenching in any background can be found in \cite{Giataganas_aniso}, but we present very briefly the method in Appendix \ref{app:jq}. It turns out that the jet quenching parameter for an energetic parton moving along the $p$ direction, while the broadening happens along the $k$ direction is given by:
\be\label{qfinal2}
\hat{q}_{p (k)}=\frac{\sqrt{2}}{\pi\a'}\left(\int_{0}^{u_h} \frac{1}{G_{kk}}\sqrt{\frac{G_{uu}}{G_{--}}}\right)^{-1}~,\quad\mbox{with}\quad
G_{--}=\frac{1}{2}(G_{00}+G_{pp})~.
\ee
It is obvious that in our anisotropic background we have three different jet quenching parameters, the number of inequivalent ways that one can place two orthogonal vectors in the anisotropic background, in contrast to the isotropic case where only one jet quenching is present. The notation we consider is summarised in the following table\newline
\begin{tabular}{|c|c|c|c|c|}
\hline
$\hat{q}$&$x_p$ & $x_k$ & Energetic parton moves along& Momentum broadening along\\\hline
$\hat{q}_{\perp(\parallel)}$&$x_\perp$ &$x_\parallel$& $x_\perp$& $x_\parallel$ \\ \hline
$\hat{q}_{\parallel(\perp)}$&$x_\parallel$ &$x_\perp$& $x_\parallel$& $x_\perp$\\
\hline
$\hat{q}_{\perp(\perp)}$&$x_{\perp,1}$ &$x_{\perp,2}$& $x_{\perp,1}$& $x_{\perp,2}$ \\
\hline
\end{tabular}\newline\newline
where the notation $x_{\parallel,\perp}$ is with respect to the anisotropic direction.

Each of the jet quenching parameters can be studied separately. When the comparison is made at equal temperatures and small anisotropies all the jet quenching parameters in presence of anisotropy are enhanced, and increase of the anisotropy leads to increase of the jet quenching parameters. The stronger enhancement happens for quarks moving along the anisotropic direction, weaker enhancement when we consider the broadening along this direction and even weaker for motion and broadening along the transverse plane
\be\label{ineqjq}
\hat{q}_{\parallel(\perp)}>\hat{q}_{\perp(\parallel)}>
\hat{q}_{\perp(\perp)}>\hat{q}_{iso}~.
\ee
We note that when the anisotropic direction is involved more actively in the calculation the deformation on the observables is stronger. This is something we have noticed in other observables too and reflects the fact that the anisotropic direction of the geometry is more heavily modified than the transverse one.

For large anisotropy the $\hat{q}_{\perp(\perp)}$  gets smaller than the isotropic parameter \cite{1203.0561}. Also the inequality between the other two parameters gets modified as $\hat{q}_{\parallel(\perp)}<\hat{q}_{\perp(\parallel)}$. Recall however that for these large anisotropies the pressure inequality \eq{pressaniso1} gets inverted and therefore the physical connection with the QGP may become more obscure.

When the comparison is made at equal entropy densities, the jet quenching parameters can be either smaller or larger than the isotropic ones depending on the values of the entropy density.

Using the singular geometry \eq{j1} it is interesting to observe that the jet quenching inequalities depend on the geometry considered \cite{1205.4684}, oblate $B=\sqrt{2}$ and prolate geometries $B=\sqrt{6}$ with the latter one giving results closer to background geometry 1.

Many different studies in the weakly coupled plasmas show that  $\hat{q}_{\perp(\parallel)}>\hat{q}_{iso}>\hat{q}_{\perp(\perp)}~$ which is partly in agreement with low anisotropy inequalities \eq{ineqjq} derived from the supergravity solution. There is also some experimental evidence supporting these results since they can explain certain asymmetric broadenings in the jet profiles \cite{star}.

In \cite{jetsu2} the jet quenching has been calculated in an unstable non-Abelian weakly coupled $SU(2)$ plasma. Considering the hard elastic collisions and soft interactions mediated by classical Yang-Mills fields, it has been found that the fields develop unstable modes with result $\hat{q}_{\perp(\parallel)}>\hat{q}_{\perp(\perp)}$. A similar relation was found in  \cite{jetqa2}. In \cite{jetqa1} the jet quenching was found to be enhanced with respect to the isotropic case, estimating it in leading logarithmic approximation by the broadening of the massless quarks interacting via gluon exchange.

\subsection{Jet Quenching of Light Probes}

In this subsection we study the jet quenching of the light probes which is related to the stopping distance of a massless particle falling in the anisotropic geometry \cite{light1,light2}. We present the simplified approach, where an induced R-charged current generated by a massless gauge field regarded as a jet traveling in the medium. We specify the distance that the jet travels in the boundary before it thermalizes, which happens when the gauge field falls into the horizon. The light quark can be introduced with a flavor D7 brane extending along the radial direction, where the infalling string in the bulk induces a flavor current on the boundary. When the string falls completely into the horizon the thermalization occurs. Null geodesics, are the ones which result the maximum stopping distances and we compute here.

It turns out that for a generic metric of the form \eq{metricqq1} with translation invariance along the 4 dimensions, the null geodesics are given by
\be
x^i\prt{u}=\int du\sqrt{G_{uu}}\frac{G^{ij}q_j}{\prt{-q_k G^{k l}q_l}^{1/2}}~,
\ee
where the indices run over the 4-dim spacetime and the $q_i$ are the constant momenta along these directions. The above relation can be simplified by choosing the space-momentum pointing along one direction ie. $q=\prt{-\omega,0,0,|\textbf{q}|}$, where $\omega$ represents the energy. Placing the probe to move along the transverse and the longitudinal direction we obtain
\be
x_\perp=\int_0^{u_h}du\prt{\frac{\omega^2}{\cB|\textbf{q}|^2}-\cF}^{-1/2}~,
\qquad
x_\parallel=\int_0^{u_h}du\prt{\frac{\cH^2 \omega^2}{\cB|\textbf{q}|^2}-\cF\cH}^{1/2}~.
\ee
The stopping distances then can be calculated in small and large anisotropies \cite{lightan}. It turns out that at small anisotropies the stopping distances decrease compared to the isotropic theory in any comparison scheme. For larger anisotropies the stopping distance along the parallel direction is smaller than the isotropic one, when keeping the temperature fixed. While along the transverse direction by fixing the entropy density the stopping distance can become bigger than the isotropic case.

\subsection{Energy Loss of Rotating Massive Quarks}

In this section we study the energy needed for a heavy probe quark to move along a circle of radius $R$ with angular frequency $\omega$, through the strongly coupled anisotropic plasma. Radiation and medium-induced energy loss playing their role in the total energy loss, and there is a crossover between a regime in which the energy loss is dominated by the drag force and a regime where the radiation from the circular motion dominates. The motivation of these studies is to understand better the radiation energy loss process \cite{radiation}.

In the anisotropic theories there are two choices of planes on which the string rotates in contrast to the isotropic ones. The simplest choice is to place the rotating string on the $SO\prt{2}$ plane and calculate the energy loss. Doing that, the progress is similar to the isotropic case since the rotational invariance is present on this plane, making the equations not so involved. A coordinate transformation along the rotational invariance plane is done to make the symmetry obvious, ie. $dx_1^2+dx_2^2\rightarrow d\rho^2+\rho^2d\phi^2$ and the ansatz for the rotating string is
\be
t=\t,\quad u=\s,\quad\rho=\rho\prt{\s},\quad\phi=\omega\t+\theta\prt{\s}~,
\ee
with the imposed condition on the boundary $\rho\prt{\infty}=R$ and $\theta\prt{\infty}=0$ in order to represent the quark cyclic motion. The functions $\rho\prt{\s}$ and $\theta\prt{\s}$ determine the spiral motion of the string hanging from the boundary. In the case where the $SO\prt{2}$ plane is chosen for rotation, the form of the string will be similar to the isotropic case. For this motion it has been argued that the energy loss due to radiation is decreased in presence of anisotropy \cite{rot1}.

To study the solution of the rotating string on the plane involving one transverse and the anisotropic direction is a more complicated problem but also interesting. There the string on the boundary may still rotate circularly, however the string in the bulk will have a deformed profile possibly close to an elliptic shape.

\subsection{Photon Production}

Since the anisotropic phase of the plasma lives for a very short time, it is essential to study probes that interact minimally with the other phases of the plasma. Such particles are the photons and dileptons produced in anisotropic phase. In strongly coupled plasmas the studies using holography were initiated and evolved in \cite{photons}.

The photon production in the field theory can be studied by a modified action of the anisotropic theory with a $U\prt{1}$ kinetic term and a relevant coupling to fundamental fields as
\be
S=S_{aniso}-\frac{1}{4}\int d^4 x\prt{F_{\m\n}^2-4 e A^\m J_\m}
\ee
with
\be
F_{\m\n}=\partial_\m A_\n-\partial_\n A_\m~,\quad J_\m=\bar{\Psi}\g_\m\Psi+\frac{i}{2}\prt{\Phi^* \cD_\m\Phi-\Phi \prt{\cD_\m\Phi}^*}~,\quad\cD_\m=D_\m-i e A_\m~.
\ee
The gravity dual of this field theory is not yet known. However a simplified approach may be applied.

To calculate the production rate of the photons in the strongly coupled anisotropic plasmas,  $N_f\ll N_c$ number of probe flavor branes representing massless quarks need to be introduced. These branes extend in the 5 dimensions of the asymptotic AdS space and wrap an $S^3\subset S^5$ of the internal space. Their field theory dual is the appearance of scalars and fermions in the fundamental representation in the quenched limit.
For the calculation of the two point correlation functions of the electromagnetic current to the leading order in $\a_{EM}$ we are interested on, it is enough to work in the anisotropic background we already know, considering only the probe branes where the dynamical photons are not present. The method used, is to  vary the string partition function with respect to $A_\m|_{u=0}$. The result in the leading order will be physically correct since the coupling of the photons to the medium is small.

It is interesting to elaborate more on the idea in the strongly coupled anisotropic plasma. In the gravity dual theory, the 4-component massless gauge field comes from the 8-dim gauge field associated to the D7 branes from the dimensional reduction to the 3-sphere. The massive Kaluza-Klein modes can be ignored in the calculation and a gauge $A_u=0$ may be taken to simplify the formulas.
By naming the remaining angles of the $S^5$, which the D7 flavor branes do not wrap, as $\theta$ and $\phi$  we consider the parametrization $\psi\prt{u}=\cos\theta$, which gives the induced metric
\bea\nn
ds^2_{D7}&=&
 \frac{1}{u^2}\prt{ -\cF \cB\, dx_0^2+dx_1^2+dx_2^2+\cH dx_3^2 +\frac{ 1-\psi^2+u^2\cF {\cal Z} \psi'^2 }{\cF\prt{1-\psi^2}}du^2}
 \\&&+ {\cal Z} \, \prt{1-\psi^2} d\Omega^2_{S^3}\,.
 \label{metricd7}
\eea
where the metric of the 5-sphere has been taken as $ds^2=d \theta^2+\cos^2\theta d\phi^2+\sin\theta^2 d\Omega_3$~. From the asymptotic behavior of $\psi$ it can be deduced that the massless quarks correspond to $\psi=0$ \cite{Mateos:2006nu}.

To get the equations of motion for the gauge field we  consider the Dirac-Born-Infeld (DBI) action
\be
S=-N_f T_{D7}\int d^8\s e^{-\phi}\sqrt{-|g+2\pi l_s^2F|},
\ee
where g is the induced metric and $T_{D7}$ is the brane tension. We expand the action and keep terms only up to quadratic level since we are interested in two-point function. The dimensional reduction to the three sphere gives the simple DBI action \cite{photonsa1}
\be
S=-\frac{N_c N_f}{16 \pi^2}\int dt d\vec{x}du\frac{e^{-3 \phi/4}\sqrt{\cB}}{u^5}F^2~.
\ee
By Fourier decomposing the components of the gauge field we get four equations of motion to solve, where one of them is decoupled from the others. To calculate the photon emission rate per unit time and volume we use the relations
\bea\label{pha1}
&&\frac{d\Gamma}{d^3\textbf{k}}=\frac{e^2}{\prt{2 \pi}^3 2 |\textbf{k}|}n_B\prt{k^0}\sum_{s=1}^2\epsilon^\m_{\prt{s}}\prt{\textbf{k}} \epsilon^\n_{\prt{s}}\prt{\textbf{k}}\chi_{\m\n}\prt{k}|_{k^0=\textbf{k}}~, ~\mbox{with}~ \chi_{\m\n}\prt{k}=-2 Im G_{\m\n}^R\prt{k}
\\
&&~\mbox{and}~ G_{\m\n}^R\prt{k}=-i\int d^4 x e^{-i k\cdot x}\Theta\prt{x_0}\vev{[J_\m^{EM}\prt{x},J_\n^{EM}\prt{0}]}~,\quad n_B\prt{k^0}=\frac{1}{e^{k^0/T}-1}~,
\eea
where $k=\prt{k^0,\textbf{k}}$ is the photon null momentum, $\chi_{\m\n}$ is the spectral density, $G_{\m\n}^R\prt{k}$ is the retarded correlator and $n_B$ is the Bose-Einstein distribution since we are in thermal equilibrium. In the anisotropic case the polarization of the vectors needs to be chosen cautiously. In \eq{pha1} each term in the summation corresponds to the number of photons emitted with polarization $\epsilon_{\prt{s}}$. To take into account the anisotropy, the vector $\textbf{k}$ can be consider in the $x_1 x_3$-plane with an angle $\varphi$ to the anisotropic $x_3$ direction. For $\varphi=0$ the momentum is chosen along the anisotropic direction and the polarization vectors lies in the transverse $SO\prt{2}$ plane. For $\varphi=\pi/2$ the momentum is chosen along the $x_1$ direction, and one polarization vector lies in anisotropic direction and the other in the transverse plane.
This choice is translated to $\textbf{k}=k_0\prt{\sin\varphi,0,\cos\varphi}$ with the orthogonal polarization vectors $\epsilon_{\prt{1}}=\prt{0,1,0}$ and $\epsilon_{\prt{2}}=\prt{\cos\varphi,0,-\sin\varphi}$~. Therefore, from \eq{pha1} we see that four different correlators need to be calculated:  $G_{2 2}^R$ for $\epsilon_{\prt{1}}$ and $G_{1 1}^R, G_{1 3}^R, G_{3 3}^R$ for $\epsilon_{\prt{2}}$.

Moreover, using these correlators the trace of the spectral function provides the electric conductivity
\be
\sigma=\frac{e^2}{4} \lim_{k^0\rightarrow 0}\frac{1}{k^0}\chi^\m_\m\prt{k}|_{k^0=\textbf{k}}~,
\ee
where the different conductivities for photons with polarization $\epsilon_{\prt{1}}$ and $\epsilon_{\prt{2}}$ normalized with the isotropic result read
\be
\sigma_{\prt{1}}= \lim_{k^0\rightarrow 0} 2 \frac{\chi_{2 2}}{\chi_{iso}}~,\quad \sigma_{\prt{2}}= \lim_{k^0\rightarrow 0} 2 \frac{\chi_{\m\n}\epsilon_{\prt{2}}^\m\epsilon_{\prt{2}}^\n}{\chi_{iso}}~.
\ee

From the results of the spectral densities it has been found that their sensitivity increases with the photon energy \cite{photonsa1}. The total production rate is always larger than the isotropic case, for all the directions of propagations. In anisotropic plasma this rate as well as the spectral densities are higher for photons with wave vectors along the longitudinal direction than the transverse one.

The conductivity for photons with polarization along the $x_2$ direction, ie.  $\epsilon_{\prt{1}}$, is independent of the direction of the wave vector and is larger than the isotropic conductivity. The conductivity for photons with polarization in the $x_1 x_3$ plane, ie.  $\epsilon_{\prt{2}}$, depends on the direction of the wave vector and along the longitudinal directions is larger than the isotropic result while in the transverse direction is smaller, irrespectively of the comparison scheme used. Notice that the photon production rate in presence of a constant magnetic field in the strongly coupled anisotropic limit, was studied in \cite{Wu:2013qja} studying the dependence of the rate to the magnetic field.

In the singular geometry 2 \eq{j1}, the photon production rate has been studied in \cite{Rebhan:2011ke}. There it was found that the DC conductivities vanish. For prolate geometries the singular geometry gives enhanced photon production rate along the anisotropic direction and suppressed along the transverse direction. For oblate geometries the results are  opposite to prolate.

Weak coupling calculations omitting the bremsstrahlung contribution, show that the photon production rate sensitivity is increased with the photon energy \cite{photon2}. In presence of anisotropy the weak coupling results suggest suppression of the photon rate production.

\section{Holographic Viscosity Bound}

The relativistic hydrodynamics describe well the QGP and simulations suggest that the QGP behaves like an almost perfect fluid \cite{0804.4015_0902.3663}. One of the most known results obtained so far from applications of gauge/gravity duality is for the prediction of the ratio of shear viscosity over entropy density \cite{etas} which was thought initially also to serve as a low bound, and only to saturate in several gauge/gravity dualities \cite{univ1,univ2}. A possible way to slightly violate that bound is the inclusion of higher derivative terms in gravities \cite{high}.  In theories with some kind of anisotropy the ratio has been studied too \cite{non} and for the p-wave superfluids some non-universal values have been obtained \cite{Erdmengera1}.

In the anisotropic theory generated by the $\theta$ space-dependent term the viscosity tensor has five components with the two independent shear viscosities being $\eta_\perp:=\eta_{x_1x_2x_1x_2}$ and $\eta_\parallel:=\eta_{x_1x_3x_1x_3}=\eta_{x_2x_3x_2x_3}.$
To obtain the desired ratios we need to take the 5-dimensional axion-dilaton gravity with the negative cosmological constant
\be
S_b=\frac{1}{2 \kappa^2}\int\sqrt{-G}\prt{R+12-\frac{\prt{\partial\phi}^2+e^{2\phi} \prt{\partial\chi}^2}{2}}~,\quad\mbox{with}\quad \kappa^2=\frac{4 \pi^2}{N_c^2}
\ee
and consider the metric fluctuations $G_{\m\n}=G_{\m\n}^\prt{0}+h_{\m\n}$ to second order, where the gauge $h_{\m u}=0$ may be taken. The expansion of the action to second order leads to a term of the form
$\int dx^5 2A\sqrt{-G}^\prt{2}$ $+\sqrt{-G}^\prt{0}\prt{R^\prt{2}-1/2 e^{2\phi} \a^2 G_{3 3}^\prt{2}}$
with \cite{etaoversa1}
\be
A=-\frac{1}{2}\prt{8+\frac{1}{2}\phi'^2 G_{uu}+\frac{1}{2}e^{2\phi}\a^2 G_{3 3}}^{\prt{0}}~.
\ee
By a Kaluza-Klein dimensional reduction along a space direction $x_i$ the above action can be mapped to an action of $U\prt{1}$ effective fields with effective coupling $g_{eff}^{-2}=G_{i i}^\prt{0}/2\kappa^2$. Notice that the effective couplings along the transverse and longitudinal directions are not equal, giving a non-universal result.

By considering the first Fourier modes $h_{2}^{1}\prt{u,q}$ and their conjugate momenta one can express the shear viscosity over entropy ratio in terms of the metric elements at the horizon of the black hole.  It has been found that the transverse shear viscosity satisfies the universal value for the isotropic Einstein gravities \cite{etaoversa1,etaoversa2}
\be
\frac{\eta_\perp}{s}=\frac{1}{4\pi}~.
\ee
This can be seen also from \cite{univ1,univ2} since the conditions of the generic proof of the universality for the ratio are satisfied for the transverse direction of our anisotropic background.

For the shear viscosity $\eta_\parallel$ the situation is more interesting since the universality conditions are not satisfied anymore. It can be found that \cite{etaoversa1,etaoversa2}
\be
\frac{\eta_\parallel}{s}=\frac{1}{4\pi\cH\prt{u_h}}=\frac{1}{4\pi} \prt{1-\frac{\a^2}{T^2} \frac{\log 2}{4\pi^2 }+\cO\prt{\a^4}}<\frac{1}{4\pi}~.
\ee
where the bound is clearly violated in the anisotropic background. Moreover,  a direct consequence of the above is that the ratio $\eta_\parallel/\eta_\perp$ is lower than the unit.

Summarizing this section, we have observed that in the anisotropic background derived by a deformation of the $\cN=4$ sYM action by a space dependent $\theta$ term, the ratio shear viscosity over entropy takes lower values than the $1/\prt{4 \pi}$ without consideration of any higher derivatives.

\section{More Quantitative Comparisons?}

In this section we attempt to obtain reasonable values for the parameters of our models aiming to make more 'quantitative' predictions of how our observables modified by the anisotropy. The  target is to specify which values of the supergravity anisotropic parameter $\a$ are sensible and physical in order to get a more 'quantitative' connection to the anisotropic QGP.

The masses of the heavy quarks are difficult to be determined in the thermal medium but for the charm and bottom quarks we have the representative values, $M_c=1.5 GeV$ and $M_b=4.8 GeV$. The medium correction to these masses can be included by specifying their dependence on the temperature and it can be done by imposing a UV regulator on the infinite straight string corresponding to the mass quark, such that the $M_Q\prt{T=0}$ is matched to the physical mass. In the anisotropic background, the medium induced corrections to the masses differ to the isotropic one due to modified horizon position, but are the same in all anisotropic directions.

The t'Hooft coupling can be fixed to $\lambda=5.5$ by comparing the static force for the conformal theory with the lattice data for relatively small separation lengths. The choice of the other parameters depend on the comparison scheme we choose, since the degrees of freedom and the field content of QCD and the anisotropic deformed theories are very different.

One comparison scheme is the fixed energy density where the energy densities of QCD and our theory are fixed and a relation between the temperatures in the two theories is obtained \cite{0902.4041}
\be
\epsilon_{QGP}\simeq\frac{\pi^2\prt{N_c^2-1+N_cN_f}}{15}T^4_{QGP}\simeq 11.2T^4_{QGP}~,\quad \epsilon_{SYM,anisot}\simeq 3 \pi^2 T^4_{SYM,anisot}~,
\ee
where the factors inside brackets in the first relation are the degrees of freedom of the $SU\prt{N_c}$ QCD, taken with three flavors and colors, while the degrees of freedom of the $\cN=4$ sYM above the phase transition has been found to be 2.7 times more than QCD $\simeq 45$. Therefore a sensible way to compare the two theories would be to associate the following relation to their temperatures at least for low anisotropic parameters $\a$
\be
T_{SYM, anisot}=2.7^{-1/4}T_{QCD}~.
\ee
Working similarly for energy densities in low anisotropy limit we get
\be
T_{SYM, anisot}=2.7^{-1/3}T_{QCD}~.
\ee
The remaining comparison scheme is to just leave the temperatures fixed and compare directly the two theories.

Now we can proceed to fix the anisotropic parameter through its naive relation to $\xi$
\be
\a^2=\frac{8\pi^2 T^2\xi}{5}~.
\ee
For small $\xi$, it  can be related to the shear viscosity of the plasma, where in one-dimensional boost invariant expansion governed by Navier-Stokes evolution \cite{0902.3834,0608270,1007.0889}
\be
\xi=\frac{10 \eta}{T\t s}~,
\ee
where the $\eta/s$ is the viscosity to entropy ratio, and $\t^{-1}$ is the expansion rate.

The dimensionless parameter $T\t$ may be estimated for the
RHIC and LHC initial conditions to
be $T\t\simeq0.35$ for RHIC conditions and $T\simeq0.43$ for LHC conditions \cite{Giataganas_aniso}. The formation times are chosen as  $\t_0\simeq 0.2 fm$, for $T=350 MeV$ and $\t_0\simeq  0.1 fm$, for $T=850 MeV$ respectively, since higher energies should give smaller initial times.

The representative value for the viscosity over entropy ratio is chosen here as $\eta/s\simeq 0.1$, and it should be noted that affects the anisotropic value significantly. The motivation for this representative choice is
that the universal prediction of AdS/CFT is $1/4\pi$. Moreover, lattice QCD simulation of a hot $SU\prt{3}$ pure hot gluon plasma at RHIC energies gives a range of values $0.1-0.4$ \cite{Nakamura:2004sy}, while a more recent result using $SU\prt{3}$ gauge theory at $T=1.165 T_c$ estimates $\eta/s\geq 0.134\prt{33}$ \cite{lns}. On the other hand in the anisotropic axion deformed geometry it is predicted from AdS/CFT that lower values of the ratio may be possible \cite{etaoversa1,etaoversa2}, but for small anisotropies the decrease is very small. Therefore, a sensible choice for the value of the ratio seems to be $\sim 0.1$, however we will examine later briefly and lower values.

Considering the above assumptions we get for the momentum space anisotropies in the initial conditions: $\xi_R\simeq2.8$ and $\xi_L\simeq 2.3$, where the smaller momentum anisotropies correspond to the LHC energies.\footnote{The indices $_R$ and $_L$ in this section correspond to the RHIC and LHC conditions respectively.}

More interesting is to consider the thermalization times,  $\t\simeq 0.6 fm$ and $T=250 MeV$ where we get the representative value $\xi_R\simeq 1.30$ for the RHIC energies. The value chosen for the thermalization time is again an issue, but we have chosen a natural representative one. Normally the hydrodynamical models in order to reproduce the magnitude of elliptic flow which observed at  RHIC, require thermalization times in the range $0.6-1 fm$. For the LHC energy, we choose as a representative value $\t\simeq 0.5 fm$ for temperatures $T=450 MeV$ which gives $\xi_L\simeq 0.87$.

In order to make the comparison, only the relation of the formation times between the theories remain to be specified. We may consider the proper times equal, or we can fix the total entropies and specify the relation of the times in the two theories, which seems to be in agreement with the discussion for the heavy quarks of \cite{qgpbook}. The formation time relation between the two theories is  approximated to
\be\label{tnormalized1}
\t_{0\cN=4~sYM}=\prt{\frac{T_{QCD}}{T_{\cN=4~sYM}}}^3
\left(\frac{d.o.f._{QCD}}{d.o.f._{\cN=4~sYM}}\right)\t_{0QCD}~,
\ee
which leads for the formation times $\t_{0\cN=4~sYM}\simeq 2.7^{-1/4} \t_{0QCD}$ and the dimensionless parameter $(T\t_0)_{\cN=4~sYM}=2.7^{-1/2} (T\t_0)_{QCD}$. Therefore close to initial conditions
\be
\xi_{\cN=4~sYM}\simeq\sqrt{2.7}~\xi_{QCD}~,
\ee
which gives for the representative values chosen $\xi_{R~\cN=4~sYM} \simeq 4.60$  and  $\xi_{L~\cN=4~sYM}\simeq 3.78$.

For comparison purposes we choose to assume that the relation between the formation times \eq{tnormalized1}, carry on approximately for the thermalization times too. In that case the values we obtain are $\xi_{R~ aSYM}\simeq 2.14$ and $\xi_{L~aSYM}\simeq 1.43$ corresponding to $(\a/T)_R\simeq 5.52$ and $(\a/T)_L\simeq 4.56$ for RHIC and LHC values respectively. This is obviously outside the region of small $\a/T$, and imply that the current background cannot be used for more quantitative predictions while the equation \eq{pressaniso1} is valid. This is true even when choosing smaller values of $\eta/s\simeq 0.05<1/4 \pi$ for the anisotropic background since then $\xi_{L~aSYM}\simeq 0.72$ and $\a/T\simeq 3.29$, where we are still far from the small anisotropic limit and in this region the pressure anisotropy gets inverted. Notice however that according to \cite{etaoversa1} the violation get increased as anisotropy increases, and for low anisotropies the $\eta/s$ is very close to $1/4\pi$. Therefore we are pushing here its values more than we actually need but still $\a/T$ is away from small anisotropic limit.
By fixing the entropy density we get  $\t_{0\cN=4~sYM}\simeq \t_{0QCD}$ and  $(T\t_0)_{\cN=4~sYM}=2.7^{-1/3} (T\t_0)_{QCD}$ resulting to
\be
\xi_{\cN=4~sYM}\simeq2.7^{1/3}~\xi_{QCD}~.
\ee The corresponding values of $\a/T$ are still large and comparable to the ones obtained with the energy density fixing. Fixing the temperatures and requiring $\t\simeq 0.7/T$ \cite{yaffetim1} for both theories, the values of $\xi$ chosen are the same in both theories $\xi_R\simeq 1.30$ and $\xi_L\simeq 0.87$, giving to $\a/T$ the values $4.36$ and $3.61$ respectively.

Therefore by following some natural assumptions and by choosing appropriate values for the parameters using different comparison schemes for the two theories, we obtain always relatively large values of $\a/T$. The physical values of anisotropy $\xi$ lead to large values of $\a/T$ where the pressure inequality in our model $P_{x_3}<P_{x_1x_2}$ get reversed and the supergravity solution does not describe anymore a moment of an expanding plasma with the desirable properties. However, in the region of small $\a/T$ the holographic plasma has the desirable properties and the observable results obtained are  qualitatively very interesting.

\section{Discussions}

In this paper we have reviewed particular strongly coupled anisotropic gauge/gravity dualities focusing mostly on the space dependent $\theta$ term theories, or equivalently from the gravity side, space dependent axion deformed theory of the original Maldacena conjecture. We have reviewed several observables and whenever possible compared with the singular background 2. The quantities that have been studied here are: \newline
$\bullet$ The static force and static potential of two heavy quarks.\newline
$\bullet$ The imaginary part of the potential. \newline
$\bullet$ The quark-antiquark pair in the plasma wind. \newline
$\bullet$ The drag force and the diffusion time for motion in the anisotropic medium. \newline
$\bullet$ The three jet quenching parameters of the anisotropic medium of a heavy moving quark. \newline
$\bullet$ The stopping distance of a light quark in the medium. \newline
$\bullet$ The energy loss of the circularly moving quark.\newline
$\bullet$ Then we have moved to study the photon production rate, an observable that is important in the anisotropic plasma since it is affected mostly from the anisotropies and interacts very weakly with the medium in the latter stages of the plasma. \newline
$\bullet$ Moreover the shear viscosity over entropy ratio was investigated and we have shown that lower values to $1/\prt{4\pi}$ may be taken in anisotropic theories.

The exact findings and methodology are reported in each section of the text. Additional new results have been included in this review, for example the drag force of the quarks in the geometry 2 and some arguments in the discussion on the validity range of the parameters in order to describe the QGP.

A quite common finding for the most observables is that when motion or measurements along the anisotropic direction happens, the expectation values are affected stronger compared to the other directions. This is a direct reflection of the degree of deformation in the anisotropic geometry where the anisotropic direction appears to be stronger deformed compared to  the transverse one.

There are several further extensions that can be considered in anisotropic gauge/gravity dualities. One of them that looks particularly interesting is the isotropization times. A way to include time dependence in the anisotropy, is to allow the axion to have a time dependence with the right initial and final conditions. By doing that the system of the supergravity equations is more involved and certain difficulties appear. Moreover, there are other ways that the anisotropy could be generated. For example the anisotropic backreaction of different dimension Dp branes may be considered. It will also be interesting to see if these new anisotropic geometries are similar to the axion deformed ones, and therefore giving qualitative similar results for the observables, establishing some consistency in the anisotropic observables. Moreover, it would be interesting if these new backgrounds allow for more physical values for their anisotropic parameters in order to achieve more 'quantitative' comparisons. \newline

\textbf{Acknowledgements:} The author would like to thank K. Sfetsos for very useful discussions. The research of the author is implemented under the "ARISTEIA" action of the "operational programme education and lifelong learning" and is co-funded by the European Social Fund (ESF) and National Resources.
 Part of this work was done while the author was supported by a SPARC and earlier by a Claude
 Leon postdoctoral fellowship. This review submitted for the Proceedings of the Corfu Summer Institute 2012
 "School and Workshops on Elementary Particle Physics and Gravity", September 8-27, 2012, Corfu, Greece.

\appendix{
\section{Q\={Q} Strings in Generic Backgrounds}\label{app:QQ}

In this section we present briefly the world-sheet calculation of a string in static gauge in gravity backgrounds which corresponds to the static potential. For a more detailed calculation the reader might look at \cite{Giataganas_aniso,stringtension}. We consider a space with metric of the form
\be\label{metricqq1}
ds^2=G_{00}d\t^2+G_{ii}dx_i^2+G_{uu}du^2~,
\ee
and choose the static gauge for the string
\be
x_0=\t\qquad\mbox{and}\qquad x_p=\sigma,
\ee
which is extended along the radial direction, so $u=u(\s)$. The $x_p$, is the direction along which the pair is aligned
\bea
x_p=x_{1,2}=:x_{\perp} \quad\mbox{or}\quad x_p=x_3=:x_{\parallel}~.\label{xpppa1}
\eea
Working in Lorentzian signature, using the Nambu-Goto action we can derive the length of the Wilson loop in the boundary in terms of $u_0$, the distance of the world-sheet in the bulk.  The length of the two endpoints of the string is given by
\be\label{staticL}
L=2\int_{u_{0}}^{\infty}  du \sqrt{\frac{- G_{uu} c^2}{(G_{00}G_{pp}+ c^2)G_{pp}}}~,
\ee
where $c$ is a constant specified by $G_{00}\prt{u_0}G_{pp}\prt{u_0}=- c^2$. The corresponding energy of the string using as renormalization method the mass subtraction of the two free quarks is
\bea\nn
2\pi\a' E= c L+2\left[  \int_{u_{0}}^{\infty} du \sqrt{- G_{uu}G_{00}}\left(\sqrt{1+\frac{c^2}{G_{pp}G_{00}}}-1\right)- \int_{u_{k}}^{u_0} du \sqrt{-G_{00}G_{uu}}\right]~,\label{staticE}
\eea
where $u_k$ is the horizon of the metric.
Therefore using \eq{staticL} and \eq{staticE} we can always find the static potential in terms of the distance of the pair.

\section{Drag Force on Trailing String in Generic Backgrounds}\label{app:drag}

We calculate the drag force analytically in a generic background, for example along the lines of \cite{kiritsis}. The analytic derivation can be found in \cite{Giataganas_aniso}. We consider the generic metric \eq{metricqq1} and the ansatz for the trailing string moving along $x_p$ direction:
\be
x_0=\t, \qquad u=\s, \qquad x_p= v \t +f(u)~,
\ee
where in the other directions the world-sheet is localized. The $x_p=x_{\parallel}$ or $x_\perp$,  as in equation \eq{xpppa1} and the function $f(u)$ at the boundary is zero in order to obtain a constant velocity quark motion.
Using the Nambu-Goto action we obtain the canonical momentum as a constant of motion constant of motion. This can be solved for $f'\prt{u}$
\be\label{dragxia11}
f'=\frac{\sqrt{-\left(G_{00}+G_{pp}v^2\right) G_{uu}}}{\sqrt{G_{00} G_{pp}\left(1+ G_{00} G_{pp}\left(2 \pi \a'\Pi_u^1 \right)^{-2}\right)}}~
\ee
and by requiring real numbers and the fact that the momentum is a constant of motion, we need to evaluate the momentum at a particular radial distance $u=u_0$ given by
\be\label{dragu0a1}
G_{uu}(G_{00} +G_{pp} v^2)=0~.
\ee
The natural solution for the drag force corresponds to the momentum flowing along the string from the boundary to the horizon. Therefore, the total drag force on the string for motion along the $x_p$ direction is
\be\label{dragfinala1}
F_{drag,x_p}=\Pi_u^1=-\sqrt{\lambda}\frac{\sqrt{-G_{00} G_{pp}}}{(2 \pi)}\bigg|_{u=u_0}~.
\ee

\section{Jet Quenching in Generic Backgrounds}\label{app:jq}

Here we present briefly the derivation of the jet quenching for generic background. The details of the full generic calculation for any background can be found in \cite{Giataganas_aniso} .\footnote{For special backgrounds which happen to satisfy $G_{--} G_{kk}=$ constant, resulting to a simplified Hamiltonian, a relation for the jet quenching was found in \cite{sfetsos2} and some extensions where studied in \cite{Armesto:2006zv}.} The light-like Wilson loop we calculate here is in the fundamental representation and the relation with the jet quenching reads \cite{jetq1}
\be\label{waf}
\vev{W^A(\cC)}\approx\exp^{-\frac{1}{4\sqrt{2}}\hat{q}L_{k}^2 L_-}~=\vev{W^F(\cC)}^2~.
\ee
To arrive to the light-cone coordinates the transformation  $\sqrt{2} x^\pm=t \pm x_p$ is done, where $x_p$ is the $x_{\parallel}$  or the $x_\perp$, as in \eq{xpppa1}~. The metric \eq{metricqq1} becomes
\bea
&&ds^2=G_{--} (dx_{+}^2+ dx_{-}^2)+G_{+-} dx_+ dx_- +G_{ii(i\neq p)}dx_i^2+G_{uu} du^2~,\\\nn
&&G_{--}=\frac{1}{2}(G_{00}+G_{pp}),\qquad G_{+-}=G_{00}-G_{pp}~.
\eea
The ansatz for the string configuration is
\bea
&&x_-=\tau,\quad x_k=\sigma,\quad u=u(\s) \\
&&x_+,~x_{i\neq p}\quad \mbox{are constant}~,
\eea
where $x_-$ has the information of the motion of the hard parton, and the other chosen direction $x_k$ is the where the small edge is aligned. Using the Nambu-Goto action we can derive the $L_k$ of the small side of the Wilson loop to be
\be
\frac{L_k}{2}=\int_{0}^{u_h} du\sqrt{\frac{c^2 G_{uu}}{(G_{kk}G_{--}-c^2)G_{kk}}}~,
\ee
where $c$ is the a constant. The length $L_k$ should be small and the above relations under certain approximations that are in general satisfied can be inverted analytically to
\be
c=\frac{L_k}{2}\left(\int_{0}^{u_h} du \frac{1}{G_{kk}}\sqrt{\frac{G_{uu}}{G_{--}}}\right)^{-1}+ \mathcal{O}(L_k^3)~.
\ee
Using the mass subtraction scheme to avoid the divergences in the total energy the normalized action becomes
\be\label{sjet1}
S=\frac{L_- L_k^2}{8\pi a'}\left(\int_{0}^{u_h} du \frac{1}{G_{kk}}\sqrt{\frac{G_{uu}}{G_{--}}}\right)^{-1} +\mathcal{O}(L_k^4)~.
\ee
Therefore the jet quenching parameter for an energetic parton moving along the $p$ direction while the broadening happens along the $k$ direction is given by:
\be\label{qfinal1}
\hat{q}_{p (k)}=\frac{\sqrt{2}}{\pi\a'}\left(\int_{0}^{u_h} \frac{1}{G_{kk}}\sqrt{\frac{G_{uu}}{G_{--}}}\right)^{-1}~.
\ee

}

\end{document}